\documentclass[12pt,twocolumn, aps,nofootinbib,preprintnumbers,superscriptaddress,numberedappendix]{openjournal}
\usepackage{natbib}

\usepackage[dvipsnames]{xcolor}
\usepackage{amsmath}
\usepackage{textgreek}
\usepackage[utf8]{inputenc}
\usepackage[english]{babel}
\usepackage[colorlinks=true, allcolors=blue]{hyperref}
\hypersetup{
    colorlinks=true,
    linkcolor=blue,
    filecolor=blue,      
    urlcolor=blue,
    citecolor=blue,
}
\usepackage{color,colortbl}

\usepackage{tensind}
\DeclareGraphicsExtensions{.bmp,.png,.jpg,.pdf}
\usepackage{verbatim}
\usepackage[normalem]{ulem}
\usepackage{orcidlink}
\usepackage{soul}
\usepackage{xspace}
\usepackage{enumitem}

\usepackage{xspace}
\usepackage{amsmath}
\usepackage[dvipsnames]{xcolor}

\newcommand\Msun{M$_{\odot}$\xspace}
\newcommand\dtdj{$\Psi_j$\xspace}

\newcommand{\orcidauthor}[3]{\author{\href{http://orcid.org/#1}{\color{blue}{#2$^{#3}$}}}}

\begin{document}

\title{\textbf{Connecting Stellar Population Surveys to Stellar Evolution with Delay-time Distributions: Application to LMC Classical Cepheids\vspace{-1.5cm}}}

\orcidauthor{0000-0002-4781-7291}{Sumit K. Sarbadhicary}{1,*}

\thanks{$^*$ Corresponding Author: \href{mailto:ssarbad1@jh.edu}{ssarbad1@jh.edu}.}

\affiliation{$^{1}$Department of Physics and Astronomy, The Johns Hopkins University, Baltimore, MD 21218 USA}

\begin{abstract}
The progenitors of many stellar-origin phenomena such as supernovae, evolved giants and supergiants, planetary nebulae, AGB stars and other post-main-sequence exotica remain poorly understood due to stellar evolution uncertainties such as mass-loss, binary evolution, rotation, mixing, and overshooting. A promising technique for investigating stellar progenitor scenarios with resolved stellar population surveys of galaxies is the delay-time distribution (DTD). Given a survey of stellar phenomena of interest, the DTD extracts the progenitor age distribution and production rates of the phenomena from star-formation history (SFH) maps of galaxies. Here we test the potential of DTDs as a stellar evolution diagnostic by applying to stars with known ages -- Classical Cepheids in the LMC. We use the high-completeness OGLE-IV survey of Cepheids, and LMC SFH maps from optical and near-infrared surveys. The measured DTDs from optical SFH maps show significant detections at ages of 20-200 Myr for FU Cepheids, and 125-200 Myr for FO Cepheids. These are consistent with independently measured ages from the period-age-color relations, with the DTD peak being more consistent with relations from non-canonical models that include effects of overshooting, rotational mixing etc. An outlier population of 0.5-0.8 Gyr of FU Cepheids is also detected in the DTD, though its veracity is debatable given the brightness and pulsation periods of FU Cepheids, and because the signal is not recovered with DTDs derived from near-infrared SFH. For upcoming surveys (e.g.\ with \emph{Roman}), DTDs of stars with well-constrained progenitors such as Cepheids and RR Lyrae can provide independent verification of the derived SFHs.
\end{abstract}

\maketitle

\section{Introduction}
\label{sec:intro}

The evolutionary pathways of many stellar  phenomena such as supernovae, planetary nebulae, asymptotic giant branch stars, evolved supergiants, X-ray binaries, and pulsational variables, remain incompletely understood. Specifically, their connection to main-sequence progenitors is complicated by uncertainties in modern stellar evolution theory, such as mass-transfer and common envelope evolution in binary systems, stellar mass-loss, and effects of convective overshooting, rotation, and chemical composition \citep[see extensive discussions in e.g][]{Iben1993,Maeder2000,Langer2012,Smith2014,Maoz2014,Eldridge2022,Marchant2024,Mathieu2025}. Stellar evolution and population synthesis models commonly incorporate these processes in parametric form, and their predictions are sensitive to the underlying parameter settings \citep{Gallart2005, Conroy2009}. Closing these gaps in our theory of stellar evolution is essential not only for stars, but also galaxies whose physical and chemical evolution is heavily influenced by stars.

A unique way to investigate progenitors of stellar  phenomena with current generation resolved stellar population surveys is the delay-time distribution or DTD \citep[see][for reviews]{Maoz2012review, Maoz2014}. Analogous to a transfer function, the DTD is the occurrence rate of a certain phenomena (e.g. SNe) versus the time elapsed (delay-time) since a delta-function burst of star-formation. If the history of star-formation and present day rates are known, the DTD for the object under consideration can be measured. These DTDs can also be predicted by stellar population synthesis simulations where models of the aforementioned uncertain physics can be varied \citep[e.g.][]{Ruiter2011, Toonen2012, Zapartas2017, Elridge2017}. A comparison between theoretical and measured DTDs can thus produce strong constraints on the physics underlying the observed phenomena. 

DTDs were originally conceived as a way to study the progenitors of Type Ia SNe \citep{Maoz2010, Maoz2011, Graur2011, Maoz2012, Graur2014, Friedmann2018, Strolger2020, Freundlich2021, Wiseman2021}. Measurements of the SN Ia DTD from different extragalactic surveys generally converged on a $t^{-1}$ form ($t$ being delay-time), providing strong support for double white dwarf progenitors for Type Ia SNe \citep{Maoz2014}. 

Given the mathematical form, DTDs can be generalized to study the progenitors of \emph{any} stellar phenomena as long as a catalog with well-quantified completeness and the associated star-formation histories (SFHs) in the survey area exist. \cite{Badenes2010} and \cite{Maoz2010snr} first demonstrated this by re-measuring the SN DTD in the LMC and SMC, using their SFH maps derived from resolved stellar photometry \citep{Harris2004,Harris2009}, and supernova remnants (SNRs) in place of SNe. \cite{Badenes2015} continued extending this technique to planetary nebulae in the LMC, showing the presence of two distinct formation channels at 35-800 Myr and at 5-8 Gyr, with the former possibly associated with a binary progenitor channel. \citet[hereafter \textbf{S21}]{Sarbadhicary2021} measured DTDs of RR Lyrae in the LMC and detected the contribution of intermediate age (1-10 Gyr) progenitors to the RR Lyrae population, in addition to the classical old ($>$10 Gyr) stellar progenitors. The presence of such intermediate-age RR Lyrae have been shown independently \citep[e.g.][]{Cuevas-Otahola2025, Mateu2025}. DTDs have also been applied to study the populations of high-mass X-ray binaries \citep{Antoniou2019} and classical novae \citep{Abelson2025}. 
The DTD technique not only makes use of full galaxy-wide surveys of these objects, but also SFH maps from resolved stellar photometry of nearby galaxies, which has seen substantial community investment over the past two decades for understanding galaxy evolution \citep[e.g][]{Harris2004, Harris2009, Weisz2013, Weisz2014, Lewis2015, Williams2017, Rubele2018, Mazzi2021, Massana2022, Cohen2024, Lazzarini2022, Mcquinn2024}.

In this paper, we present the first measurement of DTDs of Classical Cepheids in the LMC from the Optical Gravitational Lensing Experiment IV (OGLE-IV) survey \citep{ogleivofficial, sos2015a}. Classical Cepheids are pulsating supergiants with zero-age main-sequence masses of 3-13 \Msun (or ages $\sim$100 Myr) and periods of one to few hundred days \citep[see][for review]{Bono2024}. Cepheids exhibit tight correlation between their pulsation periods and luminosities \citep{Leavitt1907, Leavitt1912}, making them among the most precise tools for calibration of extragalactic distance and $H_0$ measurements \citep[e.g.][]{Hubble1929,Feast1987,Madore1991,Freedman1994, Freedman2001, Riess2019, Riess2022}.\ Cepheid periods, when combined with their mass-luminosity relations, also provide accurate, independent constraints on field stellar population ages \citep[e.g.][]{Magnier1997,Bono2005, Anderson2016, J16, Ripepi2017, DeSomma2025}. Measuring a Classical Cepheid DTD thus serves two mutually beneficial purposes: 1) Bringing a fresh perspective on Cepheid progenitor properties using resolved SFH maps, and 2) Testing how accurately DTDs can recover progenitor properties from SFH maps, given the independently measured ages of Cepheids, similar to the exercise of measuring DTDs of RR Lyrae in S21.

The paper is divided into two major sections: Section \ref{sec:methods} lays out the methodology for measuring DTDs, the two main ingredients -- Cepheid catalogs and SFH maps, and stellar evolution/pulsation models for comparison. Section \ref{sec:results} discusses the resulting DTDs and how they compare with independently measured Cepheid ages from the period-age-color relations.

\section{Measuring Delay-Time Distribution} \label{sec:methods}
Here we lay out the basic methodology for measuring DTDs, beginning with a description of key ingredients -- the catalog of Classical Cepheids and associated SFH map of the LMC (Section \ref{sec:methods:data}), and the procedure for recovering DTDs from these catalogs (Section \ref{sec:methods:dtd}). In Section \ref{sec:methods:models}, we discuss stellar evolution models and period-age relations of Cepheids for comparison with DTDs.
\subsection{Data} \label{sec:methods:data}
Measurement of DTDs for Classical Cepheids require an SFH map\footnote{\cite{Badenes2015} and S21 had also referred to these maps as `stellar age distributions' (SADs) to emphasize that these maps quantify the present day stellar mass at a given age and subregion, and not strictly the history of star-formation in that region. } and a catalog of objects for which we are measuring the DTD. 
\begin{figure*}
\includegraphics[width=\textwidth]{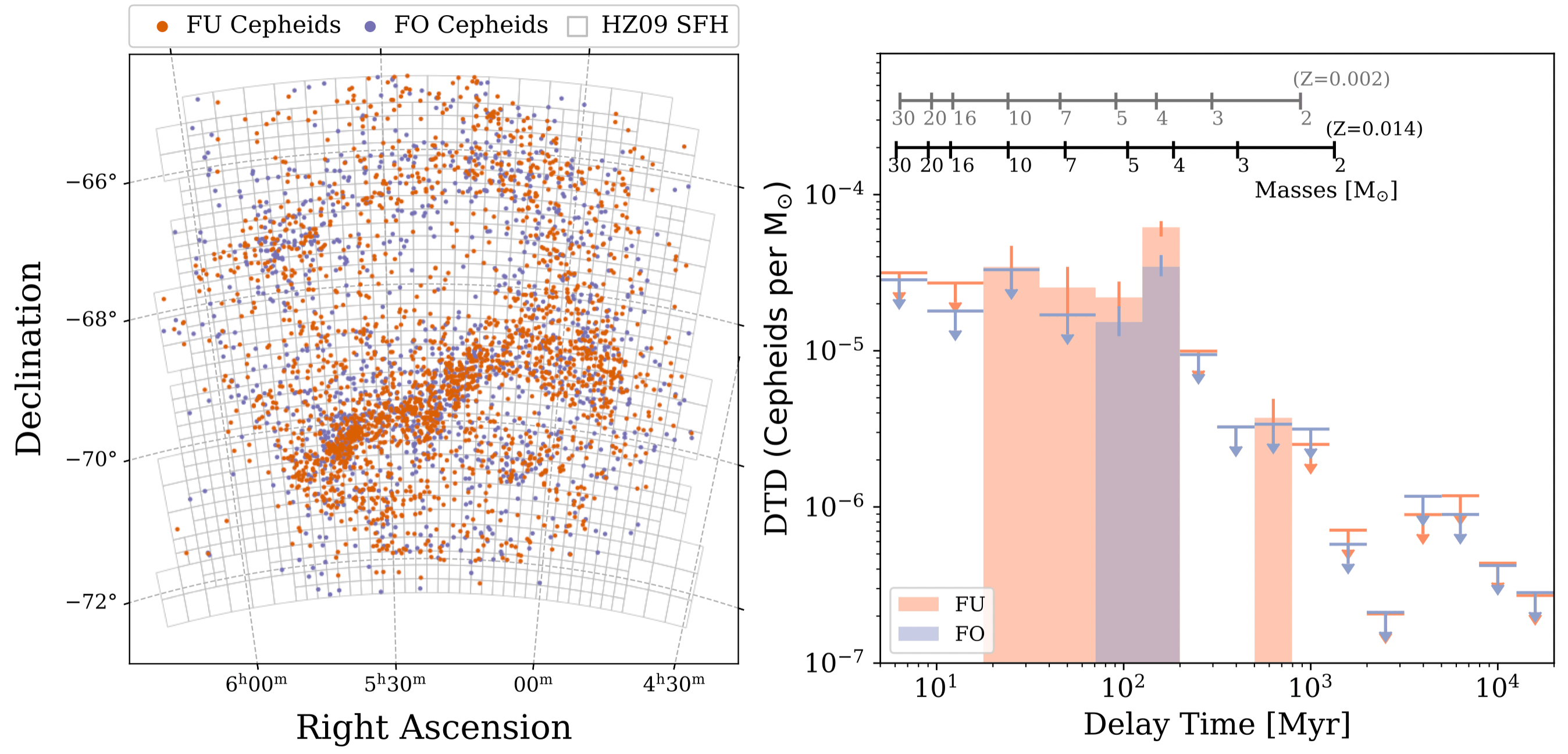}
    \caption{\textbf{Left:} Spatial distribution of fundamental (FU, orange) and first overtone (FO, purple) Cepheids from \citet[][J16]{J16}, overlaid on the stellar age distribution (SFH) map from \citet[][HZ09]{Harris2009}. \textbf{Right:} The measured DTD of the FU and FO Classical Cepheids from J16, given the SFH map from HZ09. Solid colors represent age bins where a statistically significant signal is detected in the DTD, while arrowss represent the 2$\sigma$ upper limits. Progenitor masses corresponding to the delay-times at solar and subsolar metallicity is shown with horizontal black and grey scales at the top, based on single, non-rotating stellar isochrones from PARSEC.} 
    \label{fig:cephdtdhz09}
\end{figure*}

We use the Classical Cepheid catalog in the LMC from \citet[][hereafter, \textbf{J16}]{J16}, which uses the parent sample of Cepheids from \cite{sos2015a, sos2015b} and \cite{Udalski2015}, based on time-variable $I$, $V$ band photometry from the OGLE-IV survey \citep{ogleivofficial}. A catalog of 4620 Classical Cepheids in the LMC from the OGLE-IV survey was published in \cite{sos2015a}, from which J16 removed Cepheids that deviated by more than 3$\sigma$ from the period luminosity
relation, and/or having log (period/days) $<-0.3$, as they could be affected by
blending and crowding effects. The final sample consists of 4222 Cepheids, with 2292 fundamental pulsators,
1589 first overtone pulsators. The sample completeness is
almost 99\%, based on re-detections of OGLE-III Cepheids in the OGLE-IV fields. In addition,
the OGLEIV Cepheid sample has an average $I$ = 16$\pm$0.7 mag, which is almost 6.4$\sigma$ above than the completeness limit of the most crowded field in the LMC Bar (Figure 28 in S21, $I \approx 20.5$ mag). We only include Cepheids that fall within the HZ09 SFH survey area (Figure \ref{fig:cephdtdhz09}) bringing the sample down to 2238 FU Cepheids and 1478 FO Cepehids.

J16 also provides the $V,I$ color magnitudes of the Cepheids, which will provide a useful check on the evolutionary ages of the Cepheids. We corrected these for differential reddening using $E(V-I)$ map of \cite{Skowron2021}, constructed from OGLE survey of red clump stars.

We use the SFH maps of \citet[][hereafter, \textbf{HZ09}]{Harris2009}, which was also used in S21 for measuring the LMC RR Lyrae DTD. The HZ09 SFH map contains the best-fit stellar mass vs lookback time, along with the $\pm$1$\sigma$ uncertainties, in 1376 spatial cells resolving the central 8.5\textdegree$\times$7.5\textdegree of LMC (Figure \ref{fig:cephdtdhz09}, left panel). The maps were constructed from data from the Magellanic Cloud Photometric Survey
(MCPS) of nearly 4 million stars collected with the 1 m Swope
telescope, down to a completeness of $V=20-21$ mag \citep{Zaritsky1997, Zaritsky2004}.  Each cell measures  24\arcmin$\times$24\arcmin in sky area (or 12$\times$12\arcmin if the star count exceeded 25,000), and has an SFH measured in 16 logarithmically spaced bins spanning the ages between 4 Myr and 20 Gyr, and four metallicity bins
(Z=0.008, 0.004, 0.0025, and 0.001). For ages younger than
100 Myr, a single metallicity of Z=0.008 was used as isochrones for different metallicities were indistinguishable at these ages.

We also include a discussion of how SFH measurements can affect the DTD by comparing the HZ09-based DTD with the \citet[][hereafter, \textbf{M21}]{Mazzi2021} SFH maps. The M21 SFH maps were derived using near-infrared ($YJK_s$) photometry using the 4m Visible and Near Infrared Survey Telescope for Astronomy (VISTA), conducted as part of the VISTA Survey of the Magellanic Clouds \citep[VMC,][]{Cioni2011}. The survey covers about 96 deg$^2$ of the LMC with SFH solutions measured in tiles of 0.125 deg$^2$ (or $\sim$(0.3 kpc)$^2$ projected area, assuming a distance of 50 kpc). Similar to HZ09, the M21 SFHs are derived in 16 logarithmically-spaced bins with widths of 0.3 dex for log$(t/yr)$$\leq$8.4, and 0.2 dex for older ages (slightly different from HZ09).

\begin{deluxetable*}{ll|rrr|rrr}
\tablecaption{Delay-time distributions (DTD) of Fundamental (FU) and First Overtone (FO) Cepheids. The columns $N\sigma$ and $N_{pred}$ are the detection significance of the DTD in that age bin, and the predicted number of Cepheids inside the HZ09 survey area after convolving with the SFH map, respectively. Delay-times with non-detections have values preceded by `$<$', representing the 2$\sigma$ upper limit.}
\tablehead{\multicolumn{2}{c}{} & \multicolumn{3}{c}{\textbf{FU Cepheids}} & \multicolumn{3}{c}{\textbf{FO Cepheids}}\\
\colhead{Delay Time} & \colhead{Mass} & \colhead{DTD} & \colhead{$N\sigma$} & \colhead{$N_{pred}$} & \colhead{DTD} & \colhead{$N\sigma$} & \colhead{$N_{pred}$}\\
\colhead{(Myr)} & \colhead{(M$_{\odot}$)} & \colhead{($\times$10$^{-5}$ M$^{-1}_{\odot}$)} & \colhead{} & \colhead{}& \colhead{($\times$10$^{-5}$ M$^{-1}_{\odot}$)} & \colhead{} & \colhead{}}
\startdata
$<$8 & $>$22.3 & $<$3.16 & ... & $<$67 & $<$2.85 & ... & $<$60\\
8 $-$ 17 & 12.8 $-$ 22.3 & $<$2.72 & ... & $<$117 & $<$1.80 & ... & $<$73\\
17 $-$ 35 & 8.5 $-$ 12.8 & 3.4$^{+1.3}_{-1.0}$ & 2.8 & 206$\pm$68 & $<$3.30 & ... & $<$206\\
35 $-$ 70 & 6.0 $-$ 8.5 & 2.5$^{+0.9}_{-1.0}$ & 2.7 & 199$\pm$75 & $<$1.70 & ... & $<$139\\
70 $-$ 125 & 4.6 $-$ 6.0 & 2.2$^{+0.6}_{-0.3}$ & 4.1 & 436$\pm$84 & 1.5$^{+0.4}_{-0.3}$ & 3.9 & 300$\pm$66\\
125 $-$ 200 & 3.8 $-$ 4.6 & 6.2$^{+0.6}_{-0.8}$ & 9.6 & 665$\pm$79 & 3.5$^{+0.6}_{-0.5}$ & 5.8 & 390$\pm$60\\
200 $-$ 320 & 3.1 $-$ 3.8 & $<$0.99 & ... & $<$134 & $<$0.95 & ... & $<$127\\
320 $-$ 500 & 2.6 $-$ 3.1 & $<$0.32 & ... & $<$138 & $<$0.33 & ... & $<$143\\
500 $-$ 800 & 2.2 $-$ 2.6 & 0.4$^{+0.1}_{-0.1}$ & 2.8 & 256$\pm$87 & $<$0.34 & ... & $<$249\\
800 $-$ 1260 & 1.9 $-$ 2.2 & $<$0.25 & ... & $<$106 & $<$0.32 & ... & $<$139\\
1260 $-$ 2000 & 1.5 $-$ 1.9 & $<$0.07 & ... & $<$132 & $<$0.06 & ... & $<$106\\
2000 $-$ 3160 & 1.3 $-$ 1.5 & $<$0.02 & ... & $<$60 & $<$0.02 & ... & $<$62\\
3160 $-$ 5000 & 1.1 $-$ 1.3 & $<$0.09 & ... & $<$160 & $<$0.12 & ... & $<$219\\
5000 $-$ 8000 & 1.0 $-$ 1.1 & $<$0.12 & ... & $<$292 & $<$0.09 & ... & $<$223\\
8000 $-$ 12600 & 0.9 $-$ 1.0 & $<$0.04 & ... & $<$91 & $<$0.04 & ... & $<$90\\
$>$12600& $<$0.9 & $<$0.03 & ... & $<$286 & $<$0.03 & ... & $<$309
\enddata\label{tab:cephdtd}
\end{deluxetable*}

\subsection{Methodology} \label{sec:methods:dtd}
With the catalogs above, we measured the DTD using the same methodology in S21, which we summarize here. Given an SFH map consisting of stellar mass at lookback time-bin $j$ in cell $i$ ($M_{ij}$), the DTD ($\Psi_j$), or the number of Cepheids per \Msun of stellar mass at lookback time $j$, will give the expected number of Cepheids ($\lambda_i$) in cell $i$ as 
\begin{equation} \label{eq:dtd}
\lambda_i = \sum_{i=1}^N M_{ij}\Psi_j 
\end{equation}
where $N$ is the number of age bins in the SFH map. The inverse of the above equation gives the predicted DTD ($\Psi_j$). We solve for $\Psi_j$ using a Markov Chain Monte Carlo (MCMC) inference via \texttt{emcee}, assuming a Poisson likelihood for cells with $N_i<25$ and Gaussian likelihood when $N_i\geq25$. Prior probabilities of $\Psi_j$ are assumed to be uniform in log-space, and positive valued, i.e. $\pi(\Psi_j) = 0$ for $\Psi_j\leq0$. 
\begin{figure*}
    \centering
    \includegraphics[width=\textwidth]{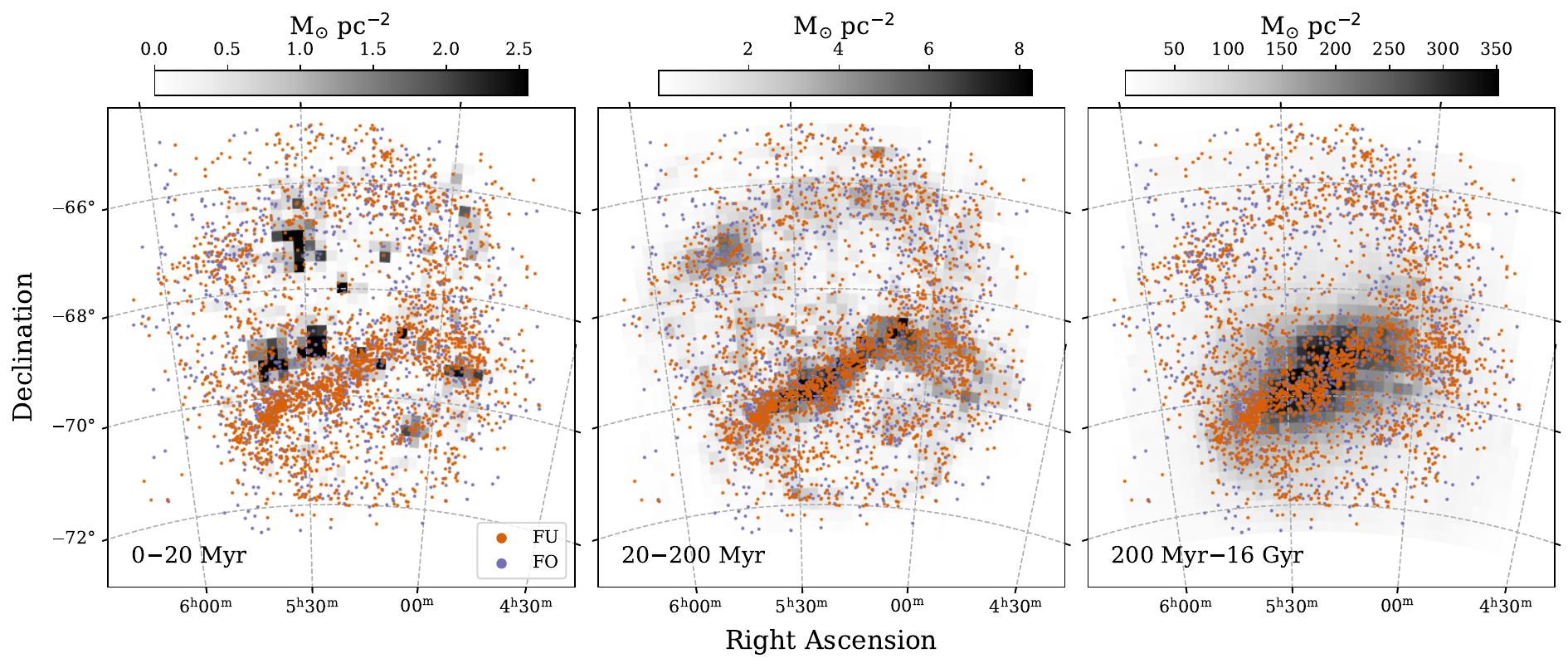}
    \caption{Visual comparison of the stellar mass surface density formed at three different lookback times in the HZ09 SFH map, with the distribution of FU+FO Cepheids from J16, both shown in orange. We show the total mass formed in three representative lookback time ranges: 0-20 Myr (left), 20-200 Myr (middle) and 200 Myr-16 Gyr (right panel). Star-formation at 20-200 Myr appears best correlated with the Cepheid distribution, in line with the DTD as shown in Figure \ref{fig:cephdtdhz09}. This age-range particularly seems to correctly account for the Cepheids in the northern and southern spiral arms. In contrast, the  youngest ($<$20 Myr) and oldest ($>$200 Myr) star-formation have excess stellar mass in cells uncorrelated with Cepheids.}
    \label{fig:cepheids_agebreakup}
\end{figure*}

The uncertainties in $\Psi_j$ include the uncertainties in the SFH using the technique introduced in S21. We create 100 different realizations of the SFH map from normal distributions centered on the median and $\pm$1$\sigma$ uncertainties of $M_{ij}$. We then measure $\Psi_j$ for each of these 100 mock SFHs, and estimate the 95\% credible interval from the combined posteriors. The mode minus the upper and lower limits of this 95\% interval is treated as our $2\sigma_+$ and $2\sigma_-$ errors respectively. We treat a given $\Psi_j$ as a ``signal'' if its value is $2\sigma_{-} > 0$. Non-detections are presented as 2$\sigma$
upper limits.

To compare our recovered Cepheid DTD with the independently measured ages of the LMC Classical Cepheids from the period-age relation (Section \ref{sec:methods:models}), we obtain the predicted number of Cepheids per lookback time bin ($\lambda_j$) by multiplying with the total stellar mass formed in each cell, i.e
\begin{equation} \label{eq:npred}
\lambda_j = \sum_{i=1}^{N_{c}} M_{ij}\Psi_j
\end{equation}
\\
representing the sum over all $N_c$ (=1376) cells of the SFH map in each lookback time bin $j$. We note that one can technically tweak this methodology e.g. by using more informative priors or combining age bins to boost signal-to-noise in specific bins. We leave this exercise for future work, and focus on demonstrating the DTD performance in the limit of no prior knowledge about progenitors, and full resolution of the SFH map.


\subsection{Predictions from Stellar Evolution Models} \label{sec:methods:models}
We will make a few points of comparison between the DTD and stellar evolution models. For a general idea of the zero-age main-sequence mass range corresponding to Cepheid delay-times, we will compare with PARSEC v2.0 isochrones for solar and sub-solar metallicity single non-rotating stars. All isochrones are derived from the publicly available \texttt{CMD}\footnote{\url{https://stev.oapd.inaf.it/cgi-bin/cmd}} interface.

In Section \ref{sec:results:parelation}, we will directly compare the number counts of Cepheids in each age bin measured by the DTD ($\lambda_j$, Eq \ref{eq:npred}) with the histogram of Cepheid ages independently measured from the period-age (PA) relation. These relations are a result of the strong mass-luminosity and period-luminosity correlation of Cepheid variables \citep{Bono2005}. Different PA relations have been explored in the literature, and as a representative example in this paper, we only consider the PA relations computed in \citet[][hereafter, \textbf{DS21}]{DS21}, using the BaSTI-IAC stellar evolution library. The PA relation is of the form $\mathrm{log}(t) = a + b(\mathrm{log}P)$, where $t$ is the age in yr, $P$ is the period in days, and $a,b$ are coefficients calibrated to the predictions from the evolution-pulsation models. With this relation, we use the measured pulsational periods given in J16 to obtain the corresponding age of each Cepheid. 

We use the values computed by DS21 for an LMC-like composition, $Z=0.008$ and $Y=0.25$, and for two different assumptions about the mass-luminosity relation: Case A, similar to \cite{Bono2005} where effects of mass-loss, rotation and overshooting are neglected, and Case B, where a luminosity offset of +0.2 dex is assumed in the mass-luminosity relation. This is expected to represent the inclusion of the above physics, which typically lead to increase of the size of the convective core during central H burning, and higher luminosities on the blue loop for a given stellar mass. From Table 6 of DS21, we use the following values for FU Cepheids: $(a,b)=(8.398, -0.776)$ for Case A, and $(a,b) = (8.503, -0.688)$ for Case B. For FO Cepheids, we use: $(a,b) = (8.28, -0.777)$, given only for Case A.

We also explore the impact of using ages from a period-age-color (PAC) relation  that adjusts for the finite width of the instability strip. The relation defined in DS21 is of the form $\mathrm{log}(t) = a + b (\mathrm{log}P) + c (V-I)$, where $V-I$ is in Johnson-Cousins filter system, applicable to OGLE-IV photometry. We use the de-reddened $V-I$ colors obtained in Section \ref{sec:methods:data}. From Table 7 of DS21, we use the following values for FU Cepheids: $(a,b,c) = (8.728, -0.345, -0.794)$ for Case A, and $(a,b,c) = (8.6, -0.581, -0.216)$ for Case B. For FO Cepheids, DS21 only provides relation for Case A in their Table 8: $(a,b,c) = (8.04, -0.599, 0.208)$.

\section{Results and Discussion} \label{sec:results}
With the DTD methodology defined in Section \ref{sec:methods}, we present the measured DTDs for the FU and FO Cepheids in Section \ref{sec:results:dtd}, and discuss how they compare with independently measured Cepheid ages from the period-age-color relations in Section \ref{sec:results:parelation}. 

\subsection{The measured Cepheid DTD} \label{sec:results:dtd}
Figure \ref{fig:cephdtdhz09} shows the DTD obtained with the HZ09 SFH map, with values tabulated in Table \ref{tab:cephdtd}.\ The right panel of Figure \ref{fig:cephdtdhz09} shows the FU Cepheid DTD, which has significant signal at 20-200 Myr, corresponding to zero-age main-sequence masses of $\sim$3.8$-$12.8 \Msun at LMC metallicity for single stars. The production rate of Cepheids in this age range is  about $(2-6) \times 10^{-5}$ Cepheids per \Msun of stars. A solitary contribution from the 0.5-0.8 Gyr age bin is also seen in Figure \ref{fig:cephdtdhz09}, corresponding to $\sim$2.2$-$2.6 \Msun progenitors (which we discuss later in Section \ref{sec:results:0.5-0.8gyr}). The FO Cepheids have significant detections in their DTD between 70-200 Myr, or $\sim$3.6$-$6 \Msun progenitors. The strongest detection is in the 125-200 Myr bins for both FU (at 9.6$\sigma$) and FO (at 5.8$\sigma$), and this age bin accounts for about 29$\pm$4\% and 25$\pm$5\% of the FU and FO Cepheids, respectively, according to the DTD. Altogether, the DTD at 20-200 Myr and 500-800 Myr bins can account for about 76$\pm$6 \% of the FU Cepheids, while the 70-200 Myr bin accounts for about 45$\pm$7 \% of the FO Cepheids. The remaining age bins may account for the remainder of the Cepheids, but are not associated with statistically significant signals above 0 from the MCMC chains, and are hence represented as 2$\sigma$ upper limits.
\begin{figure*}
\centering
    \includegraphics[width=\textwidth]{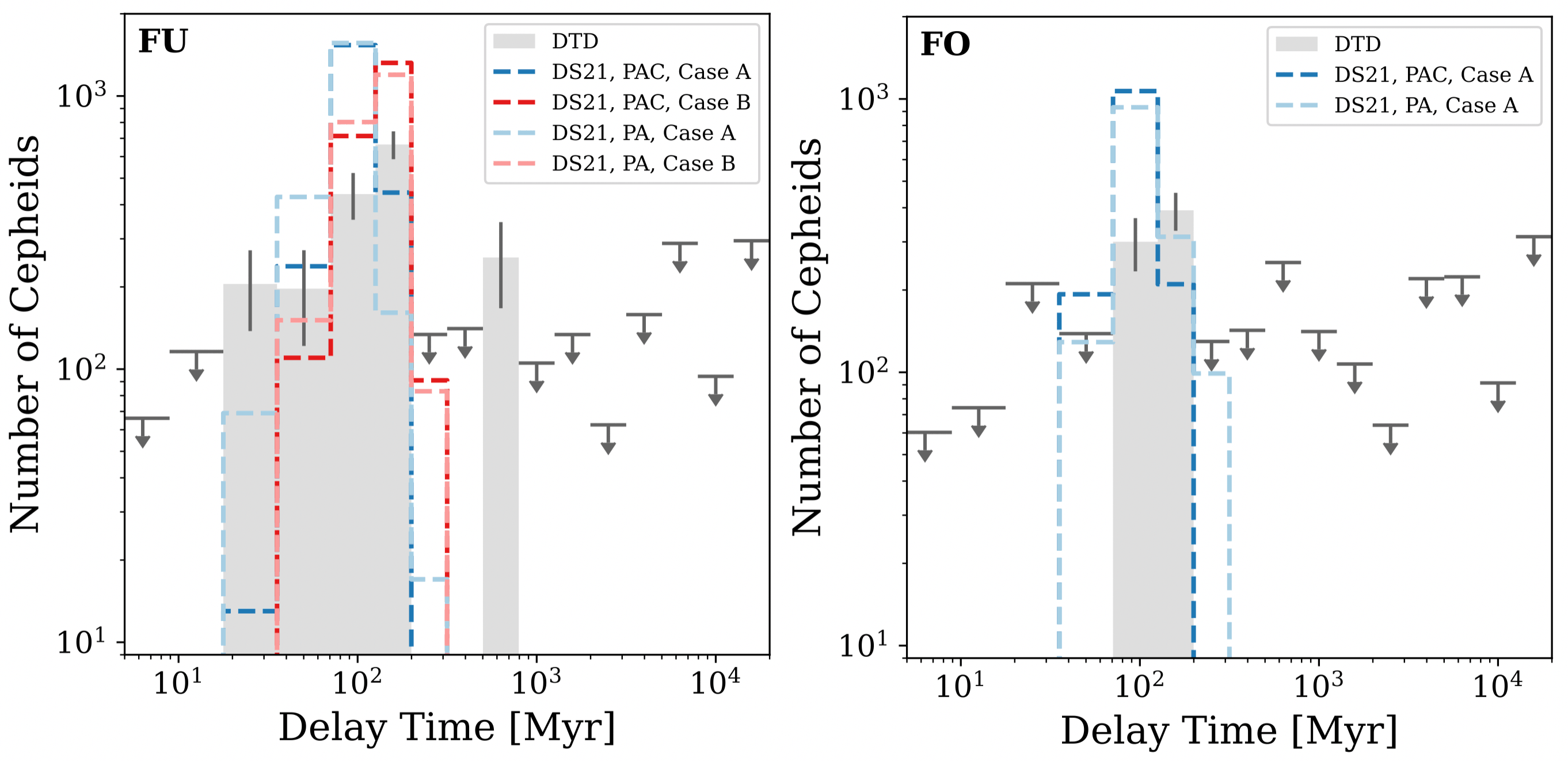}
    \caption{Comparison between the DTD and the independently measured ages of FU (\textbf{left}) and FO (\textbf{right}) Cepheids from the theoretical period-age (PA) and period-age color (PAC) relations of \citet[][DS21]{DS21}. The grey histograms show the DTD-based age distribution of Cepheids (grey), obtained from convolving the DTDs in Figure \ref{fig:cephdtdhz09} with the total stellar mass formed in each cell and age bin. The dashed histograms show the PA-based (lighter color) and PAC-based (darker color) age distributions from DS21. Blue histograms represent relations in DS21 obtained by neglecting overshooting, mass-loss and rotation (Case A), while red histograms account for these effects with an offset mass-luminosity relation (Case B). See Section \ref{sec:methods:models} for more details.}  
    \label{fig:compwithobs}
\end{figure*}

The measured DTD appears to be consistent with the spatial correlation of the Cepheids and the SFH map in Figure \ref{fig:cepheids_agebreakup}. The amount and distribution of stellar mass in the 20-200 Myr bin appears consistent with the numbers and distribution of Cepheids, including the populations in the bar, the northern and southern spiral arms. In contrast, the stellar mass at much younger ($<$20 Myr) and older ($>$0.2 Gyr) ages have less spatial consistency. Stellar mass formed at ages$<$20 Myr are concentrated in the young HII regions of LMC, where only a small number of Cepheids are located. The stellar mass formed in the older age bin show a more uniform distribution that significantly peaks along the central bar, which does not account for the full observed structure of the Cepheid population. 
\begin{figure}
    \centering
    \includegraphics[width=\columnwidth]{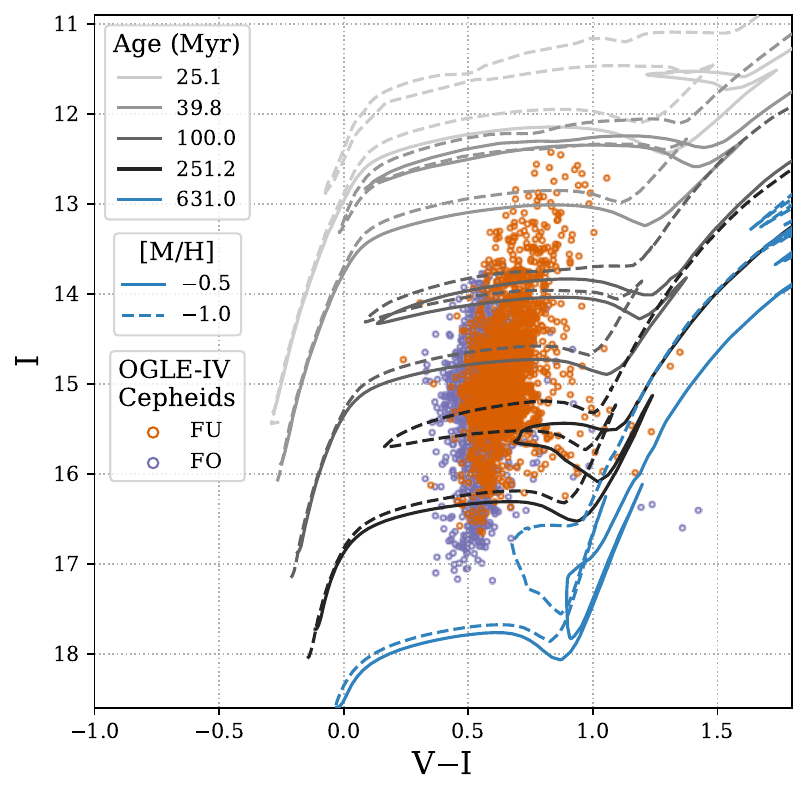}
    \caption{Mean reddening-corrected $I$ vs $V-I$ photometry of FU (brown) and FO (purple) Cepheids, and comparison with PARSEC v2.0 isochrones for different ages and two metallicities, [M/H]=[$-$0.5, $-$1.0], covering the lower range expected in the LMC. The comparison here broadly validates the ages being recovered by the DTDs in Figures \ref{fig:cephdtdhz09}-\ref{fig:compwithobs}, though see discussion in Section \ref{sec:results}.}
    \label{fig:cepheidcmd}
\end{figure}

\subsection{Comparison with Period-Age-Color Relations} \label{sec:results:parelation}
We compare the measured DTDs of the FU and FO Cepheids with their independently measured ages from the PA and PAC relations in Figure \ref{fig:compwithobs}. The PA and PAC-based ages are derived according to Section \ref{sec:methods:models}, while the DTD-based ages were obtained from Eq \ref{eq:npred} in Section \ref{sec:methods:dtd}, by convolving the DTD with the stellar mass formed in each age bin of the HZ09 map.

From a glance at Figure \ref{fig:compwithobs}, we find that, barring the 0.5-0.8 Gyr signal for the FU (which we discuss in Section \ref{sec:results:0.5-0.8gyr}), the DTD-based age range is consistent with the different PA and PAC-based age-ranges, with values roughly between 20-300 Myr for the FU Cepheids and about 40-300 Myr for FO Cepheids. Some difference in the predicted age range PA-based and PAC-based relations is noticeable, with the PAC producing a narrower age range the Cepheids. 

The age range also appears to be consistent with Figure \ref{fig:cepheidcmd}, where the low-metallicity PARSEC isochrones in a similar age range overlap with the median $V$ and $I$ color magnitude positions of the Cepheids, particularly on the first crossings and blue loops. This overall consistency is encouraging and lends support to the DTD as a progenitor diagnostic, considering that it is effectively recovering all this age information from just the positions of Cepheids with respect to the underlying SFH map. 

Although the ages implied by the DTD and the PA/PAC relations are similar, the shapes of the age distributions are somewhat different. Both Case A and Case B PA/PAC relations appear to indicate more Cepheids with ages of 100 Myrs than implied by the DTD, as well as under-predict Cepheids in the youngest detected bin (17-35 Myr) of the DTD. We note a caveat here regarding the comparison -- while the PA/PAC relations provide precise ages for each individual Cepheid, and thus well-determined number of Cepheids per age-bin, the DTD does not. The number of Cepheids per \Msun in each age bin of the DTD are free-parameters measured statistically (Section \ref{sec:methods:dtd}), and thus by definition, some fraction of the sample is locked up in the upper limits. 

Figure \ref{fig:compwithobs} does however seem to suggest that the DTD is more consistent with the Case B age distribution than Case A. The Case B ages are older, leading to the peak of the distribution at $\sim$150 Myr, which is more aligned with the maxima of the DTD. In contrast, Case A predicts younger ages, peaking at $\sim$100 Myr for FU and FO Cepheids.  As a reminder, Case A models ignore the effects of overshooting, rotation and mass-loss in the derivation of PA and PAC relations (Section \ref{sec:methods:models}). The result is consistent with independent findings that dynamical and pulsation-based masses of Cepheids converge more with models that include moderate amounts of overshooting, rotation and/or mass-loss \citep[e.g][]{Pietrzynski2010,PradaMornoni2012, Marconi2013,Ragosta2019}.
 \begin{figure*}
    \centering
    \includegraphics[width=\textwidth]{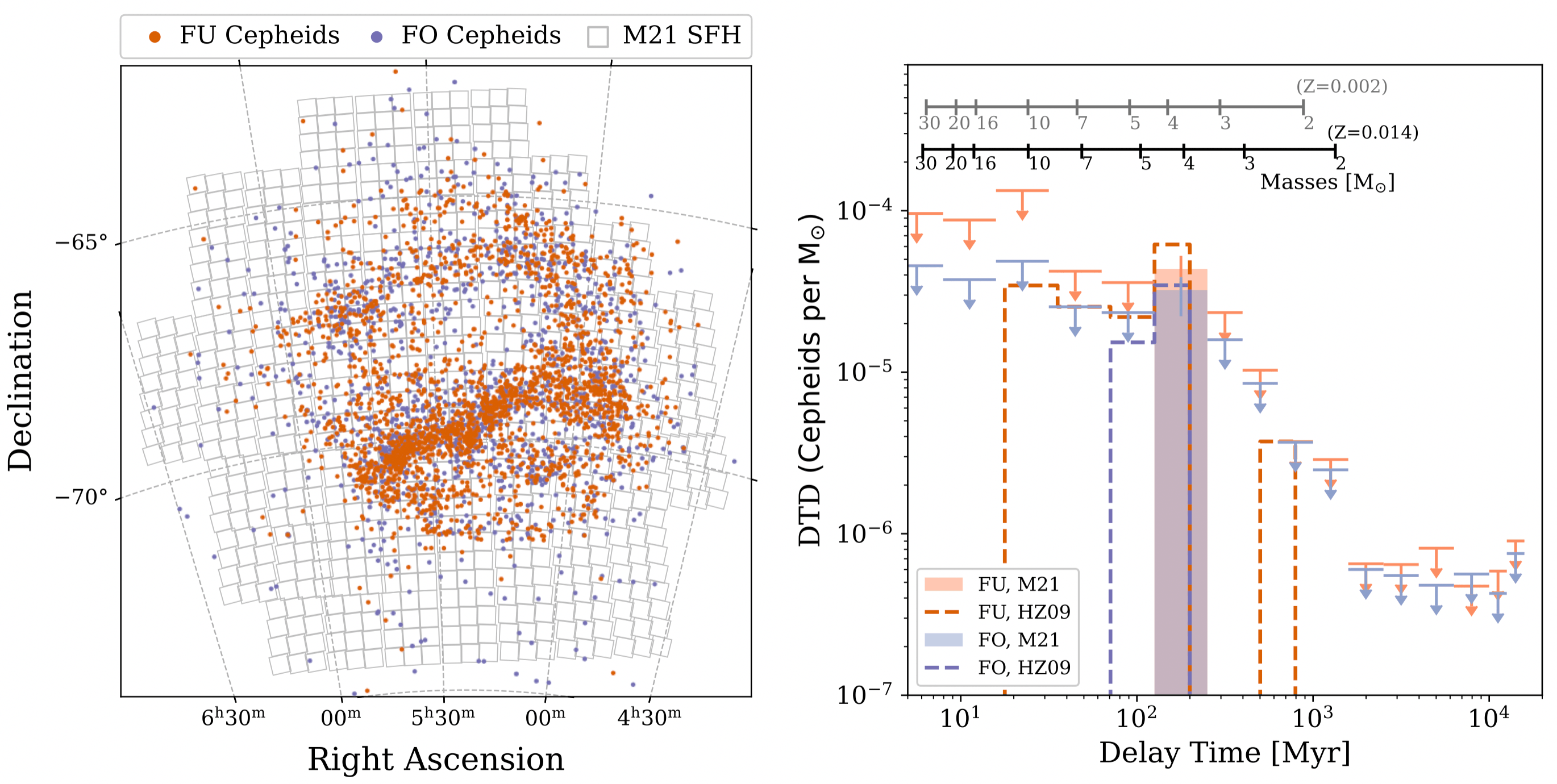}
    \caption{\textbf{Left:} The J16 Classical Cepheid population in the LMC shown on the \citet[][referred to as M21]{Mazzi2021} SFH map, with spatial cells shown in grey. \textbf{Right:} The DTD for FU and FO Cepheids derived from the M21 SFH map (filled histogram + upper limits) and comparison with the HZ09-based DTDs from Figure \ref{fig:cephdtdhz09} (solid dashed).}
    \label{fig:cepheidm21}
\end{figure*}

\subsubsection{Are there Classical FU Cepheids with ages of 0.5-0.8 Gyr?} \label{sec:results:0.5-0.8gyr}
The most notable discrepancy between the DTD and the PA/PAC relations in Figure \ref{fig:cephdtdhz09} is the apparent population of FU Cepheids at 0.5-0.8 Gyr. The signal in this bin is weak $\sim$2.8 $\sigma$, but above our defined 2$\sigma$ threshold. We carried out a two checks introduced in S21 to ensure the validity of the signal: 
\begin{enumerate}
    \item We checked if this signal is a false positive using the artificial DTD test done in Section 4.1 of S21. Specifically, we repeat the FU DTD measurement procedure in Section \ref{sec:methods:dtd} but with mock Cepheid maps generated from a FU DTD that is identical to Figure \ref{fig:cephdtdhz09} but excluding the 0.5-0.8 Gyr signal (i.e \dtdj in this bin is set to 0). We find that the recovered DTD from this mock Cepheid map \emph{does not} have a signal at 0.5-0.8 Gyr. Based on this test, it is unlikely that the observed 0.5-0.8 Gyr signal recovered in the HZ09 DTD (Figure \ref{fig:cephdtdhz09}) is a false positive, but rather driven by an apparent correlation between some fraction of the OGLE-IV Cepheids and the measured SFH at lookback times of 0.5-0.8 Gyr in the LMC.
    \item We also checked if the shallow crowding-limited photometry in the LMC Bar could be driving an erroneous 0.5-0.8 Gyr detection. In this case, we recalculated the FU Cepheid DTD by excluding cells belonging to the Inner and Outer Bars as defined in HZ09 (see Section 4.2 and Figures 6,7 in S21). We find that the remeasured DTD without the Bar cells still contain the 0.5-0.8 Gyr signal at a significance of about $\sim$2.7$\sigma$. This implies that the signal is not localized to just the Bar, but to the full HZ09 map.
\end{enumerate} 

If real, the 0.5-0.8 Gyr signal implies that about 10\% of the FU Cepheids in OGLE IV have masses 2.2-2.6 \Msun, smaller than the typically quoted lower limit of Classical Cepheids at $\sim$3 \Msun. On one hand, the presence of these low-mass Cepheids appears inconsistent with their color-magnitude properties (Figure \ref{fig:cepheidcmd}). With respect to the 0.6 Gyr isochrone, the FU Cepheids are about $I\sim1$ mag brighter than the first crossing, and $V-I \sim 0.2-0.5$ mag bluer than the He-burning blue loop\footnote{We do note that the extent of the blue loop is heavily dependent on assumptions in stellar evolution models. While our comparison is with the specific PARSEC isochrones, the point to note is that the $\lesssim$200 Myr isochrones produce blue loops that overlap with the J16 Cepheids, while the 600 Myr isochrone does not.}. On the other hand, these isochrones do assume single-stellar evolution. Short-period Cepheids down to $\sim$3 \Msun have been found in binary systems, and interpreted as Cepheids on the first crossing through the instability strip during H-shell burning phase \citep[e.g][]{BohmVitense1998, Pilecki2022}. A large fraction ($>$50\%) of Cepheids are expected in binaries \citep{Kervella2019}, and interactions are known to dramatically alter the course of stars from their single-stellar tracks \citep{Neilson2015, Karczmarek2022}. A similar signal is not observed in the fainter FO Cepheids, though their smaller sample could be a factor.  

Given the inconsistencies above, we  consider an alternative explanation -- that systematic uncertainties in the SFH solutions could be driving the 0.5-0.8 Gyr signal. A similar explanation was posited in S21 regarding the detection of intermediate-aged ($1-10$ Gyr) RR Lyrae with the HZ09 SFH map. While the existence of RR Lyrae at these ages now seems plausible from independent studies \citep[e.g][]{Iorio2021, Cuevas-Otahola2025, Mateu2025}, the large fractional contribution ($\gtrsim$50\%) at these ages implied by the DTD is in tension with the metal enrichment history of the LMC (see discussion in Section 5 of S21).

We test the above hypothesis by measuring the Cepheid DTD with a different SFH, namely from M21 which was derived from near-infrared $YJK_s$ observations. We refer the reader again to Section \ref{sec:methods:data} for the basic details of the M21 SFH. The DTD was re-measured using the same procedure as HZ09 laid out in Section \ref{sec:methods:dtd}.

The recovered M21-based DTD and its comparison with the HZ09-based DTD is shown in Figure \ref{fig:cepheidm21}. The M21 DTD has a single detection in the age bin 125-250 Myr, coincident with the peak of the HZ09 DTD, and upper limits for the other bins. The 0.5-0.8 Gyr signal that was detected in HZ09 however, is not detected with the M21 SFH at the 2$\sigma$ level. On the other hand, the M21 DTD for FU Cepheids does not show detections at younger ages (20$-$100 Myr), though the 2$\sigma$ upper limits at these ages are consistent with detections in HZ09. 

This indicates that the measured DTD can be sensitive to the fidelity of the spatially-resolved SFH solutions. M21 had noted that their SFHs under-estimate the stellar mass formed at ages $\lesssim$100 Myr compared to HZ09 in most sub-regions of LMC (see the discussion in Sections 5.2-5.4 of M21). M21 attributed these discrepancies to differences in survey strategies with HZ09, such as photometric bands used ($UBVI$ in HZ09 which are more sensitive to younger stellar ages than $YJK_s$ in M21), and the methodologies for SFH derivation, such as assumptions about the initial mass function and relation between age and metallicity. More recently, \cite{Abelson2025} showed with their  classical novae DTD computed with the SFH map of M31 \citep{Williams2017} that the choice of stellar isochrone model for SFH construction can also have a sizeable impact on the recovered DTDs.

\section{SUMMARY}

We tested how accurately the method of delay-time distribution (DTD) can recover information about progenitors of a stellar-origin phenomena by applying it to Classical Cepheids in the LMC. Cepheids have the unique advantage of accurate, independent measurements of ages from their pulsational periods that can verify the recovered DTDs.\ We calculate DTDs of fundamental (FU) and first-overtone (FO) Cepheids from the OGLE-IV survey compiled in \cite{J16}, using star formation history (SFH) maps derived from optical photometry in \citet{Harris2009} and near-infrared (NIR) photometry in \citet{Mazzi2021}. We measured DTDs following the Bayesian methodology in \cite{Sarbadhicary2021}, accounting for both statistical and SFH errors. We summarize our findings as follows:

\begin{itemize}
\item \textit{Ages of Cepheids}: The DTDs were broadly consistent with the known age range of Classical Cepheids, showing significant detections at 20-200 Myr for FU Cepheids ($\sim$66\% of the FU sample), and 125-200 Myr for FO Cepheids ($\sim$45\% of the sample). The corresponding progenitor mass range of 3.8$-$12.8 \Msun is consistent with the typically quoted range in the literature.

\item \textit{Production rate of Cepheids:} Based on the measured DTDs, FU Cepheids in the detected age-range are produced at the rate of $2.2-6.6$ per 10$^5$ \Msun of stars, while FO Cepheids are produced at $1.5-3.5$ per 10$^5$ \Msun.  
   
\item \textit{Consistency with period-age-color relations:} The DTDs appear to be mostly consistent with the various period age color relations of Cepheids. The peak of the DTD-based age distribution (at 125-200 Myr) is consistent with period-age relations derived from non-canonical models, i.e.\ ones that do not ignore overshooting, rotational, and mass-loss.

\item \textit{Presence of old FU Cepheids?} The FU DTD suggests the presence of a 0.5-0.8 Gyr population, accounting for about 10\% of the sample, and notably discrepant with the period age relations. The population is detected in the DTD measured with the \cite{Harris2009} SFH map, but not with the \cite{Mazzi2021} SFH map, highlighting the potential impact of systematic uncertainties in the SFH solutions on the recovered DTD.
\end{itemize}

The results demonstrate that DTDs are an effective tool for extracting information about stellar progenitors of unknown phenomena using resolved stellar population surveys of nearby galaxies. With an arguably complete catalog of Cepheids overlaid on the SFH map of the LMC, we were able to identify the relevant age range and contribution of different progenitors to the observed population of LMC Cepheids. The accuracy of the progenitor information recovered by the DTDs however is contingent on the accuracy of spatially-resolved SFH solutions, which can be affected by various factors such as photometric coverage, differences in SFH construction methodology, and choice of stellar isochrone models. We therefore also advocate for the use of DTDs as a validation step for SFH measurements with upcoming high-resolution wide-field photometric surveys from \emph{Roman}, \emph{Rubin} and \emph{Euclid}. As shown here and in S21, DTDs of variable stars such as RR Lyrae and Cepheids, which have independently constrained ages and large, high-completeness catalogs, can provide particularly useful and independent sanity checks on the measured SFHs.



\section*{Acknowledgments}

SKS acknowledges useful discussion, inspiration for the project,  and years of support from his PhD thesis advisor, Carles Badenes. SKS also thanks the OGLE, MCPS and VMC collaborations for making their data public and easily accessible for analysis.

\bibliographystyle{aasjournal}
\bibliography{main}

@ARTICLE{Mathieu2025,
       author = {{Mathieu}, Robert D. and {Pols}, Onno R.},
        title = "{Blue Stragglers and Friends: Initial Evolutionary Pathways in Close Low-Mass Binaries}",
      journal = {\araa},
         year = 2025,
        month = aug,
       volume = {63},
       number = {1},
        pages = {467-512},
          doi = {10.1146/annurev-astro-071221-054402},
archivePrefix = {arXiv},
       eprint = {2509.20531},
 primaryClass = {astro-ph.SR},
       adsurl = {https://ui.adsabs.harvard.edu/abs/2025ARA&A..63..467M}
}

@ARTICLE{Marchant2024,
       author = {{Marchant}, Pablo and {Bodensteiner}, Julia},
        title = "{The Evolution of Massive Binary Stars}",
      journal = {\araa},
         year = 2024,
        month = sep,
       volume = {62},
       number = {1},
        pages = {21-61},
          doi = {10.1146/annurev-astro-052722-105936},
archivePrefix = {arXiv},
       eprint = {2311.01865},
 primaryClass = {astro-ph.SR},
       adsurl = {https://ui.adsabs.harvard.edu/abs/2024ARA&A..62...21M}
}

@ARTICLE{Smith2014,
       author = {{Smith}, Nathan},
        title = "{Mass Loss: Its Effect on the Evolution and Fate of High-Mass Stars}",
      journal = {\araa},
         year = 2014,
        month = aug,
       volume = {52},
        pages = {487-528},
          doi = {10.1146/annurev-astro-081913-040025},
archivePrefix = {arXiv},
       eprint = {1402.1237},
 primaryClass = {astro-ph.SR},
       adsurl = {https://ui.adsabs.harvard.edu/abs/2014ARA&A..52..487S}
}

@ARTICLE{Langer2012,
       author = {{Langer}, N.},
        title = "{Presupernova Evolution of Massive Single and Binary Stars}",
      journal = {\araa},
         year = 2012,
        month = sep,
       volume = {50},
        pages = {107-164},
          doi = {10.1146/annurev-astro-081811-125534},
archivePrefix = {arXiv},
       eprint = {1206.5443},
 primaryClass = {astro-ph.SR},
       adsurl = {https://ui.adsabs.harvard.edu/abs/2012ARA&A..50..107L}
}

@ARTICLE{Gallart2005,
       author = {{Gallart}, C. and {Zoccali}, M. and {Aparicio}, A.},
        title = "{The Adequacy of Stellar Evolution Models for the Interpretation of the Color-Magnitude Diagrams of Resolved Stellar Populations}",
      journal = {\araa},
         year = 2005,
        month = sep,
       volume = {43},
       number = {1},
        pages = {387-434},
          doi = {10.1146/annurev.astro.43.072103.150608},
       adsurl = {https://ui.adsabs.harvard.edu/abs/2005ARA&A..43..387G}
}

@ARTICLE{Conroy2009,
       author = {{Conroy}, Charlie and {Gunn}, James E. and {White}, Martin},
        title = "{The Propagation of Uncertainties in Stellar Population Synthesis Modeling. I. The Relevance of Uncertain Aspects of Stellar Evolution and the Initial Mass Function to the Derived Physical Properties of Galaxies}",
      journal = {\apj},
         year = 2009,
        month = jul,
       volume = {699},
       number = {1},
        pages = {486-506},
          doi = {10.1088/0004-637X/699/1/486},
archivePrefix = {arXiv},
       eprint = {0809.4261},
 primaryClass = {astro-ph},
       adsurl = {https://ui.adsabs.harvard.edu/abs/2009ApJ...699..486C}
}

@ARTICLE{Maeder2000,
       author = {{Maeder}, Andr{\'e} and {Meynet}, Georges},
        title = "{The Evolution of Rotating Stars}",
      journal = {\araa},
         year = 2000,
        month = jan,
       volume = {38},
        pages = {143-190},
          doi = {10.1146/annurev.astro.38.1.143},
archivePrefix = {arXiv},
       eprint = {astro-ph/0004204},
 primaryClass = {astro-ph},
       adsurl = {https://ui.adsabs.harvard.edu/abs/2000ARA&A..38..143M}
}

@ARTICLE{Iben1993,
       author = {{Iben}, Jr., Icko and {Livio}, Mario},
        title = "{Common Envelopes in Binary Star Evolution}",
      journal = {\pasp},
         year = 1993,
        month = dec,
       volume = {105},
        pages = {1373},
          doi = {10.1086/133321},
       adsurl = {https://ui.adsabs.harvard.edu/abs/1993PASP..105.1373I}
}

@ARTICLE{Eldridge2022,
       author = {{Eldridge}, Jan J. and {Stanway}, Elizabeth R.},
        title = "{New Insights into the Evolution of Massive Stars and Their Effects on Our Understanding of Early Galaxies}",
      journal = {\araa},
         year = 2022,
        month = aug,
       volume = {60},
        pages = {455-494},
          doi = {10.1146/annurev-astro-052920-100646},
archivePrefix = {arXiv},
       eprint = {2202.01413},
 primaryClass = {astro-ph.GA},
       adsurl = {https://ui.adsabs.harvard.edu/abs/2022ARA&A..60..455E}
}

@ARTICLE{Mazzi2021,
       author = {{Mazzi}, Alessandro and {Girardi}, L{\'e}o and {Zaggia}, Simone and {Pastorelli}, Giada and {Rubele}, Stefano and {Bressan}, Alessandro and {Cioni}, Maria-Rosa L. and {Clementini}, Gisella and {Cusano}, Felice and {Rocha}, Jo{\~a}o Pedro and {Gullieuszik}, Marco and {Kerber}, Leandro and {Marigo}, Paola and {Ripepi}, Vincenzo and {Bekki}, Kenji and {Bell}, Cameron P.~M. and {de Grijs}, Richard and {Groenewegen}, Martin A.~T. and {Ivanov}, Valentin D. and {Oliveira}, Joana M. and {Sun}, Ning-Chen and {van Loon}, Jacco Th},
        title = "{The VMC survey - XLIII. The spatially resolved star formation history across the Large Magellanic Cloud}",
      journal = {\mnras},
         year = 2021,
        month = nov,
       volume = {508},
       number = {1},
        pages = {245-266},
          doi = {10.1093/mnras/stab2399},
archivePrefix = {arXiv},
       eprint = {2108.07225},
 primaryClass = {astro-ph.GA},
       adsurl = {https://ui.adsabs.harvard.edu/abs/2021MNRAS.508..245M}
}

@ARTICLE{Maoz2010snr,
       author = {{Maoz}, Dan and {Badenes}, Carles},
        title = "{The supernova rate and delay time distribution in the Magellanic Clouds}",
      journal = {\mnras},
         year = 2010,
        month = sep,
       volume = {407},
       number = {2},
        pages = {1314-1327},
          doi = {10.1111/j.1365-2966.2010.16988.x},
archivePrefix = {arXiv},
       eprint = {1003.3031},
 primaryClass = {astro-ph.GA},
       adsurl = {https://ui.adsabs.harvard.edu/abs/2010MNRAS.407.1314M}
}

@ARTICLE{Mcquinn2024,
       author = {{McQuinn}, Kristen B.~W. and {Newman}, Max J.~B. and {Skillman}, Evan D. and {Telford}, O. Grace and {Brooks}, Alyson and {Adams}, Elizabeth A.~K. and {Berg}, Danielle A. and {Boyer}, Martha L. and {Cannon}, John M. and {Dolphin}, Andrew E. and {Pahl}, Anthony J. and {Rhode}, Katherine L. and {Salzer}, John J. and {Cohen}, Roger E. and {Goldman}, Steve R.},
        title = "{The Ancient Star Formation History of the Extremely Low-mass Galaxy Leo P: An Emerging Trend of a Post-reionization Pause in Star Formation}",
      journal = {\apj},
         year = 2024,
        month = nov,
       volume = {976},
       number = {1},
          eid = {60},
        pages = {60},
          doi = {10.3847/1538-4357/ad8158},
archivePrefix = {arXiv},
       eprint = {2409.19050},
 primaryClass = {astro-ph.GA},
       adsurl = {https://ui.adsabs.harvard.edu/abs/2024ApJ...976...60M}
}

@ARTICLE{Weisz2014,
       author = {{Weisz}, Daniel R. and {Dolphin}, Andrew E. and {Skillman}, Evan D. and {Holtzman}, Jon and {Gilbert}, Karoline M. and {Dalcanton}, Julianne J. and {Williams}, Benjamin F.},
        title = "{The Star Formation Histories of Local Group Dwarf Galaxies. I. Hubble Space Telescope/Wide Field Planetary Camera 2 Observations}",
      journal = {\apj},
         year = 2014,
        month = jul,
       volume = {789},
       number = {2},
          eid = {147},
        pages = {147},
          doi = {10.1088/0004-637X/789/2/147},
archivePrefix = {arXiv},
       eprint = {1404.7144},
 primaryClass = {astro-ph.GA},
       adsurl = {https://ui.adsabs.harvard.edu/abs/2014ApJ...789..147W}
}

@ARTICLE{Lazzarini2022,
       author = {{Lazzarini}, Margaret and {Williams}, Benjamin F. and {Durbin}, Meredith J. and {Dalcanton}, Julianne J. and {Smercina}, Adam and {Bell}, Eric F. and {Choi}, Yumi and {Dolphin}, Andrew and {Gilbert}, Karoline and {Guhathakurta}, Puragra and {Rosolowsky}, Erik and {Skillman}, Evan and {Telford}, O. Grace and {Weisz}, Daniel},
        title = "{The Panchromatic Hubble Andromeda Treasury: Triangulum Extended Region (PHATTER). II. The Spatially Resolved Recent Star Formation History of M33}",
      journal = {\apj},
         year = 2022,
        month = jul,
       volume = {934},
       number = {1},
          eid = {76},
        pages = {76},
          doi = {10.3847/1538-4357/ac7568},
archivePrefix = {arXiv},
       eprint = {2206.11393},
 primaryClass = {astro-ph.GA},
       adsurl = {https://ui.adsabs.harvard.edu/abs/2022ApJ...934...76L}
}

@ARTICLE{Lewis2015,
       author = {{Lewis}, Alexia R. and {Dolphin}, Andrew E. and {Dalcanton}, Julianne J. and {Weisz}, Daniel R. and {Williams}, Benjamin F. and {Bell}, Eric F. and {Seth}, Anil C. and {Simones}, Jacob E. and {Skillman}, Evan D. and {Choi}, Yumi and {Fouesneau}, Morgan and {Guhathakurta}, Puragra and {Johnson}, Lent C. and {Kalirai}, Jason S. and {Leroy}, Adam K. and {Monachesi}, Antonela and {Rix}, Hans-Walter and {Schruba}, Andreas},
        title = "{The Panchromatic Hubble Andromeda Treasury. XI. The Spatially Resolved Recent Star Formation History of M31}",
      journal = {\apj},
         year = 2015,
        month = jun,
       volume = {805},
       number = {2},
          eid = {183},
        pages = {183},
          doi = {10.1088/0004-637X/805/2/183},
archivePrefix = {arXiv},
       eprint = {1504.03338},
 primaryClass = {astro-ph.GA},
       adsurl = {https://ui.adsabs.harvard.edu/abs/2015ApJ...805..183L}
}

@ARTICLE{Williams2017,
       author = {{Williams}, Benjamin F. and {Dolphin}, Andrew E. and {Dalcanton}, Julianne J. and {Weisz}, Daniel R. and {Bell}, Eric F. and {Lewis}, Alexia R. and {Rosenfield}, Philip and {Choi}, Yumi and {Skillman}, Evan and {Monachesi}, Antonela},
        title = "{PHAT. XIX. The Ancient Star Formation History of the M31 Disk}",
      journal = {\apj},
         year = 2017,
        month = sep,
       volume = {846},
       number = {2},
          eid = {145},
        pages = {145},
          doi = {10.3847/1538-4357/aa862a},
       adsurl = {https://ui.adsabs.harvard.edu/abs/2017ApJ...846..145W}
}

@ARTICLE{Cohen2024,
       author = {{Cohen}, Roger E. and {McQuinn}, Kristen B.~W. and {Murray}, Claire E. and {Williams}, Benjamin F. and {Choi}, Yumi and {Lindberg}, Christina W. and {Burhenne}, Clare and {Gordon}, Karl D. and {Yanchulova Merica-Jones}, Petia and {Gilbert}, Karoline M. and {Boyer}, Martha L. and {Goldman}, Steven and {Dolphin}, Andrew E. and {Telford}, O. Grace},
        title = "{Scylla. II. The Spatially Resolved Star Formation History of the Large Magellanic Cloud Reveals an Inverted Radial Age Gradient}",
      journal = {\apj},
         year = 2024,
        month = nov,
       volume = {975},
       number = {1},
          eid = {42},
        pages = {42},
          doi = {10.3847/1538-4357/ad6cd5},
archivePrefix = {arXiv},
       eprint = {2410.11696},
 primaryClass = {astro-ph.GA},
       adsurl = {https://ui.adsabs.harvard.edu/abs/2024ApJ...975...42C}
}

@ARTICLE{Massana2022,
       author = {{Massana}, P. and {Ruiz-Lara}, T. and {No{\"e}l}, N.~E.~D. and {Gallart}, C. and {Nidever}, D.~L. and {Choi}, Y. and {Sakowska}, J.~D. and {Besla}, G. and {Olsen}, K.~A.~G. and {Monelli}, M. and {Dorta}, A. and {Stringfellow}, G.~S. and {Cassisi}, S. and {Bernard}, E.~J. and {Zaritsky}, D. and {Cioni}, M.-R.~L. and {Monachesi}, A. and {van der Marel}, R.~P. and {de Boer}, T.~J.~L. and {Walker}, A.~R.},
        title = "{The synchronized dance of the magellanic clouds' star formation history}",
      journal = {\mnras},
         year = 2022,
        month = jun,
       volume = {513},
       number = {1},
        pages = {L40-L45},
          doi = {10.1093/mnrasl/slac030},
archivePrefix = {arXiv},
       eprint = {2203.09523},
 primaryClass = {astro-ph.GA},
       adsurl = {https://ui.adsabs.harvard.edu/abs/2022MNRAS.513L..40M}
}

@ARTICLE{Rubele2018,
       author = {{Rubele}, Stefano and {Pastorelli}, Giada and {Girardi}, L{\'e}o and {Cioni}, Maria-Rosa L. and {Zaggia}, Simone and {Marigo}, Paola and {Bekki}, Kenji and {Bressan}, Alessandro and {Clementini}, Gisella and {de Grijs}, Richard and {Emerson}, Jim and {Groenewegen}, Martin A.~T. and {Ivanov}, Valentin D. and {Muraveva}, Tatiana and {Nanni}, Ambra and {Oliveira}, Joana M. and {Ripepi}, Vincenzo and {Sun}, Ning-Chen and {van Loon}, Jacco Th},
        title = "{The VMC survey - XXXI: The spatially resolved star formation history of the main body of the Small Magellanic Cloud}",
      journal = {\mnras},
         year = 2018,
        month = aug,
       volume = {478},
       number = {4},
        pages = {5017-5036},
          doi = {10.1093/mnras/sty1279},
archivePrefix = {arXiv},
       eprint = {1805.04516},
 primaryClass = {astro-ph.GA},
       adsurl = {https://ui.adsabs.harvard.edu/abs/2018MNRAS.478.5017R}
}

@ARTICLE{Weisz2013,
       author = {{Weisz}, Daniel R. and {Dolphin}, Andrew E. and {Skillman}, Evan D. and {Holtzman}, Jon and {Dalcanton}, Julianne J. and {Cole}, Andrew A. and {Neary}, Kyle},
        title = "{Comparing the ancient star formation histories of the Magellanic Clouds}",
      journal = {\mnras},
         year = 2013,
        month = may,
       volume = {431},
       number = {1},
        pages = {364-371},
          doi = {10.1093/mnras/stt165},
archivePrefix = {arXiv},
       eprint = {1301.7422},
 primaryClass = {astro-ph.CO},
       adsurl = {https://ui.adsabs.harvard.edu/abs/2013MNRAS.431..364W}
}

@ARTICLE{Harris2004,
       author = {{Harris}, Jason and {Zaritsky}, Dennis},
        title = "{The Star Formation History of the Small Magellanic Cloud}",
      journal = {\aj},
         year = 2004,
        month = mar,
       volume = {127},
       number = {3},
        pages = {1531-1544},
          doi = {10.1086/381953},
archivePrefix = {arXiv},
       eprint = {astro-ph/0312100},
 primaryClass = {astro-ph},
       adsurl = {https://ui.adsabs.harvard.edu/abs/2004AJ....127.1531H}
}

@ARTICLE{Antoniou2019,
       author = {{Antoniou}, Vallia and {Zezas}, Andreas and {Drake}, Jeremy J. and {Badenes}, Carles and {Haberl}, Frank and {Wright}, Nicholas J. and {Hong}, Jaesub and {Di Stefano}, Rosanne and {Gaetz}, Terrance J. and {Long}, Knox S. and {Plucinsky}, Paul P. and {Sasaki}, Manami and {Williams}, Benjamin F. and {Winkler}, P. Frank and {SMC XVP Collaboration}},
        title = "{Deep Chandra Survey of the Small Magellanic Cloud. III. Formation Efficiency of High-mass X-Ray Binaries}",
      journal = {\apj},
         year = 2019,
        month = dec,
       volume = {887},
       number = {1},
          eid = {20},
        pages = {20},
          doi = {10.3847/1538-4357/ab4a7a},
archivePrefix = {arXiv},
       eprint = {1901.01237},
 primaryClass = {astro-ph.HE},
       adsurl = {https://ui.adsabs.harvard.edu/abs/2019ApJ...887...20A}
}

@ARTICLE{Badenes2010,
       author = {{Badenes}, Carles and {Maoz}, Dan and {Draine}, Bruce T.},
        title = "{On the size distribution of supernova remnants in the Magellanic Clouds}",
      journal = {\mnras},
         year = 2010,
        month = sep,
       volume = {407},
       number = {2},
        pages = {1301-1313},
          doi = {10.1111/j.1365-2966.2010.17023.x},
archivePrefix = {arXiv},
       eprint = {1003.3030},
 primaryClass = {astro-ph.GA},
       adsurl = {https://ui.adsabs.harvard.edu/abs/2010MNRAS.407.1301B}
}

@ARTICLE{Maoz2012review,
       author = {{Maoz}, D. and {Mannucci}, F.},
        title = "{Type-Ia Supernova Rates and the Progenitor Problem: A Review}",
      journal = {\pasa},
         year = 2012,
        month = jan,
       volume = {29},
       number = {4},
        pages = {447-465},
          doi = {10.1071/AS11052},
archivePrefix = {arXiv},
       eprint = {1111.4492},
 primaryClass = {astro-ph.CO},
       adsurl = {https://ui.adsabs.harvard.edu/abs/2012PASA...29..447M}
}

@ARTICLE{Maoz2012,
       author = {{Maoz}, Dan and {Mannucci}, Filippo and {Brandt}, Timothy D.},
        title = "{The delay-time distribution of Type Ia supernovae from Sloan II}",
      journal = {\mnras},
         year = 2012,
        month = nov,
       volume = {426},
       number = {4},
        pages = {3282-3294},
          doi = {10.1111/j.1365-2966.2012.21871.x},
archivePrefix = {arXiv},
       eprint = {1206.0465},
 primaryClass = {astro-ph.CO},
       adsurl = {https://ui.adsabs.harvard.edu/abs/2012MNRAS.426.3282M}
}

@ARTICLE{Strolger2020,
       author = {{Strolger}, Louis-Gregory and {Rodney}, Steven A. and {Pacifici}, Camilla and {Narayan}, Gautham and {Graur}, Or},
        title = "{Delay Time Distributions of Type Ia Supernovae from Galaxy and Cosmic Star Formation Histories}",
      journal = {\apj},
         year = 2020,
        month = feb,
       volume = {890},
       number = {2},
          eid = {140},
        pages = {140},
          doi = {10.3847/1538-4357/ab6a97},
archivePrefix = {arXiv},
       eprint = {2001.05967},
 primaryClass = {astro-ph.GA},
       adsurl = {https://ui.adsabs.harvard.edu/abs/2020ApJ...890..140S}
}

@ARTICLE{Wiseman2021,
       author = {{Wiseman}, P. and {Sullivan}, M. and {Smith}, M. and {Frohmaier}, C. and {Vincenzi}, M. and {Graur}, O. and {Popovic}, B. and {Armstrong}, P. and {Brout}, D. and {Davis}, T.~M. and {Galbany}, L. and {Hinton}, S.~R. and {Kelsey}, L. and {Kessler}, R. and {Lidman}, C. and {M{\"o}ller}, A. and {Nichol}, R.~C. and {Rose}, B. and {Scolnic}, D. and {Toy}, M. and {Zontou}, Z. and {Asorey}, J. and {Carollo}, D. and {Glazebrook}, K. and {Lewis}, G.~F. and {Tucker}, B.~E. and {Abbott}, T.~M.~C. and {Aguena}, M. and {Allam}, S. and {Andrade-Oliveira}, F. and {Annis}, J. and {Bacon}, D. and {Bertin}, E. and {Brooks}, D. and {Buckley-Geer}, E. and {Burke}, D.~L. and {Carnero Rosell}, A. and {Carrasco Kind}, M. and {Carretero}, J. and {Costanzi}, M. and {da Costa}, L.~N. and {Pereira}, M.~E.~S. and {Desai}, S. and {Diehl}, H.~T. and {Doel}, P. and {Everett}, S. and {Ferrero}, I. and {Flaugher}, B. and {Fosalba}, P. and {Frieman}, J. and {Garc{\'\i}a-Bellido}, J. and {Gaztanaga}, E. and {Giannantonio}, T. and {Gruen}, D. and {Gruendl}, R.~A. and {Gschwend}, J. and {Gutierrez}, G. and {Hollowood}, D.~L. and {Honscheid}, K. and {Hoyle}, B. and {James}, D.~J. and {Krause}, E. and {Kuehn}, K. and {Kuropatkin}, N. and {Maia}, M.~A.~G. and {Marshall}, J.~L. and {Martini}, P. and {Menanteau}, F. and {Miquel}, R. and {Morgan}, R. and {Ogando}, R.~L.~C. and {Palmese}, A. and {Paz-Chinch{\'o}n}, F. and {Petravick}, D. and {Pieres}, A. and {Plazas Malag{\'o}n}, A.~A. and {Romer}, A.~K. and {Sanchez}, E. and {Scarpine}, V. and {Schubnell}, M. and {Serrano}, S. and {Sevilla-Noarbe}, I. and {Soares-Santos}, M. and {Suchyta}, E. and {Swanson}, M.~E.~C. and {Tarle}, G. and {Thomas}, D. and {To}, C. and {Varga}, T.~N. and {Walker}, A.~R. and {DES Collaboration}},
        title = "{Rates and delay times of Type Ia supernovae in the Dark Energy Survey}",
      journal = {\mnras},
         year = 2021,
        month = sep,
       volume = {506},
       number = {3},
        pages = {3330-3348},
          doi = {10.1093/mnras/stab1943},
archivePrefix = {arXiv},
       eprint = {2105.11954},
 primaryClass = {astro-ph.GA},
       adsurl = {https://ui.adsabs.harvard.edu/abs/2021MNRAS.506.3330W}
}

@ARTICLE{Freundlich2021,
       author = {{Freundlich}, Jonathan and {Maoz}, Dan},
        title = "{The delay time distribution of Type-Ia supernovae in galaxy clusters: the impact of extended star-formation histories}",
      journal = {\mnras},
         year = 2021,
        month = apr,
       volume = {502},
       number = {4},
        pages = {5882-5895},
          doi = {10.1093/mnras/stab493},
archivePrefix = {arXiv},
       eprint = {2012.00793},
 primaryClass = {astro-ph.GA},
       adsurl = {https://ui.adsabs.harvard.edu/abs/2021MNRAS.502.5882F}
}

@ARTICLE{Friedmann2018,
       author = {{Friedmann}, Matan and {Maoz}, Dan},
        title = "{The rate of Type-Ia supernovae in galaxy clusters and the delay-time distribution out to redshift 1.75}",
      journal = {\mnras},
         year = 2018,
        month = sep,
       volume = {479},
       number = {3},
        pages = {3563-3581},
          doi = {10.1093/mnras/sty1664},
archivePrefix = {arXiv},
       eprint = {1803.04421},
 primaryClass = {astro-ph.GA},
       adsurl = {https://ui.adsabs.harvard.edu/abs/2018MNRAS.479.3563F}
}

@ARTICLE{Graur2011,
       author = {{Graur}, O. and {Poznanski}, D. and {Maoz}, D. and {Yasuda}, N. and {Totani}, T. and {Fukugita}, M. and {Filippenko}, A.~V. and {Foley}, R.~J. and {Silverman}, J.~M. and {Gal-Yam}, A. and {Horesh}, A. and {Jannuzi}, B.~T.},
        title = "{Supernovae in the Subaru Deep Field: the rate and delay-time distribution of Type Ia supernovae out to redshift 2}",
      journal = {\mnras},
         year = 2011,
        month = oct,
       volume = {417},
       number = {2},
        pages = {916-940},
          doi = {10.1111/j.1365-2966.2011.19287.x},
archivePrefix = {arXiv},
       eprint = {1102.0005},
 primaryClass = {astro-ph.CO},
       adsurl = {https://ui.adsabs.harvard.edu/abs/2011MNRAS.417..916G}
}

@ARTICLE{Graur2014,
       author = {{Graur}, O. and {Rodney}, S.~A. and {Maoz}, D. and {Riess}, A.~G. and {Jha}, S.~W. and {Postman}, M. and {Dahlen}, T. and {Holoien}, T.~W.-S. and {McCully}, C. and {Patel}, B. and {Strolger}, L.-G. and {Ben{\'\i}tez}, N. and {Coe}, D. and {Jouvel}, S. and {Medezinski}, E. and {Molino}, A. and {Nonino}, M. and {Bradley}, L. and {Koekemoer}, A. and {Balestra}, I. and {Cenko}, S.~B. and {Clubb}, K.~I. and {Dickinson}, M.~E. and {Filippenko}, A.~V. and {Frederiksen}, T.~F. and {Garnavich}, P. and {Hjorth}, J. and {Jones}, D.~O. and {Leibundgut}, B. and {Matheson}, T. and {Mobasher}, B. and {Rosati}, P. and {Silverman}, J.~M. and {U}, V. and {Jedruszczuk}, K. and {Li}, C. and {Lin}, K. and {Mirmelstein}, M. and {Neustadt}, J. and {Ovadia}, A. and {Rogers}, E.~H.},
        title = "{Type-Ia Supernova Rates to Redshift 2.4 from CLASH: The Cluster Lensing And Supernova Survey with Hubble}",
      journal = {\apj},
         year = 2014,
        month = mar,
       volume = {783},
       number = {1},
          eid = {28},
        pages = {28},
          doi = {10.1088/0004-637X/783/1/28},
archivePrefix = {arXiv},
       eprint = {1310.3495},
 primaryClass = {astro-ph.CO},
       adsurl = {https://ui.adsabs.harvard.edu/abs/2014ApJ...783...28G}
}

@ARTICLE{Maoz2011,
       author = {{Maoz}, Dan and {Mannucci}, Filippo and {Li}, Weidong and {Filippenko}, Alexei V. and {Della Valle}, Massimo and {Panagia}, Nino},
        title = "{Nearby supernova rates from the Lick Observatory Supernova Search - IV. A recovery method for the delay-time distribution}",
      journal = {\mnras},
         year = 2011,
        month = apr,
       volume = {412},
       number = {3},
        pages = {1508-1521},
          doi = {10.1111/j.1365-2966.2010.16808.x},
archivePrefix = {arXiv},
       eprint = {1002.3056},
 primaryClass = {astro-ph.CO},
       adsurl = {https://ui.adsabs.harvard.edu/abs/2011MNRAS.412.1508M}
}

@ARTICLE{Maoz2010,
       author = {{Maoz}, Dan and {Sharon}, Keren and {Gal-Yam}, Avishay},
        title = "{The Supernova Delay Time Distribution in Galaxy Clusters and Implications for Type-Ia Progenitors and Metal Enrichment}",
      journal = {\apj},
         year = 2010,
        month = oct,
       volume = {722},
       number = {2},
        pages = {1879-1894},
          doi = {10.1088/0004-637X/722/2/1879},
archivePrefix = {arXiv},
       eprint = {1006.3576},
 primaryClass = {astro-ph.CO},
       adsurl = {https://ui.adsabs.harvard.edu/abs/2010ApJ...722.1879M}
}

@ARTICLE{Elridge2017,
       author = {{Eldridge}, J.~J. and {Stanway}, E.~R. and {Xiao}, L. and {McClelland}, L.~A.~S. and {Taylor}, G. and {Ng}, M. and {Greis}, S.~M.~L. and {Bray}, J.~C.},
        title = "{Binary Population and Spectral Synthesis Version 2.1: Construction, Observational Verification, and New Results}",
      journal = {\pasa},
         year = 2017,
        month = nov,
       volume = {34},
          eid = {e058},
        pages = {e058},
          doi = {10.1017/pasa.2017.51},
archivePrefix = {arXiv},
       eprint = {1710.02154},
 primaryClass = {astro-ph.SR},
       adsurl = {https://ui.adsabs.harvard.edu/abs/2017PASA...34...58E}
}

@ARTICLE{Zapartas2017,
       author = {{Zapartas}, E. and {de Mink}, S.~E. and {Izzard}, R.~G. and {Yoon}, S.-C. and {Badenes}, C. and {G{\"o}tberg}, Y. and {de Koter}, A. and {Neijssel}, C.~J. and {Renzo}, M. and {Schootemeijer}, A. and {Shrotriya}, T.~S.},
        title = "{Delay-time distribution of core-collapse supernovae with late events resulting from binary interaction}",
      journal = {\aap},
         year = 2017,
        month = may,
       volume = {601},
          eid = {A29},
        pages = {A29},
          doi = {10.1051/0004-6361/201629685},
archivePrefix = {arXiv},
       eprint = {1701.07032},
 primaryClass = {astro-ph.HE},
       adsurl = {https://ui.adsabs.harvard.edu/abs/2017A&A...601A..29Z}
}

@ARTICLE{Ruiter2011,
       author = {{Ruiter}, A.~J. and {Belczynski}, K. and {Sim}, S.~A. and {Hillebrandt}, W. and {Fryer}, C.~L. and {Fink}, M. and {Kromer}, M.},
        title = "{Delay times and rates for Type Ia supernovae and thermonuclear explosions from double-detonation sub-Chandrasekhar mass models}",
      journal = {\mnras},
         year = 2011,
        month = oct,
       volume = {417},
       number = {1},
        pages = {408-419},
          doi = {10.1111/j.1365-2966.2011.19276.x},
archivePrefix = {arXiv},
       eprint = {1011.1407},
 primaryClass = {astro-ph.SR},
       adsurl = {https://ui.adsabs.harvard.edu/abs/2011MNRAS.417..408R}
}

@ARTICLE{Toonen2012,
       author = {{Toonen}, S. and {Nelemans}, G. and {Portegies Zwart}, S.},
        title = "{Supernova Type Ia progenitors from merging double white dwarfs. Using a new population synthesis model}",
      journal = {\aap},
         year = 2012,
        month = oct,
       volume = {546},
          eid = {A70},
        pages = {A70},
          doi = {10.1051/0004-6361/201218966},
archivePrefix = {arXiv},
       eprint = {1208.6446},
 primaryClass = {astro-ph.HE},
       adsurl = {https://ui.adsabs.harvard.edu/abs/2012A&A...546A..70T}
}

@ARTICLE{Maoz2014,
       author = {{Maoz}, Dan and {Mannucci}, Filippo and {Nelemans}, Gijs},
        title = "{Observational Clues to the Progenitors of Type Ia Supernovae}",
      journal = {\araa},
         year = 2014,
        month = aug,
       volume = {52},
        pages = {107-170},
          doi = {10.1146/annurev-astro-082812-141031},
archivePrefix = {arXiv},
       eprint = {1312.0628},
 primaryClass = {astro-ph.CO},
       adsurl = {https://ui.adsabs.harvard.edu/abs/2014ARA&A..52..107M}
}

@ARTICLE{Cioni2011,
       author = {{Cioni}, M.-R.~L. and {Clementini}, G. and {Girardi}, L. and {Guandalini}, R. and {Gullieuszik}, M. and {Miszalski}, B. and {Moretti}, M.-I. and {Ripepi}, V. and {Rubele}, S. and {Bagheri}, G. and {Bekki}, K. and {Cross}, N. and {de Blok}, W.~J.~G. and {de Grijs}, R. and {Emerson}, J.~P. and {Evans}, C.~J. and {Gibson}, B. and {Gonzales-Solares}, E. and {Groenewegen}, M.~A.~T. and {Irwin}, M. and {Ivanov}, V.~D. and {Lewis}, J. and {Marconi}, M. and {Marquette}, J.-B. and {Mastropietro}, C. and {Moore}, B. and {Napiwotzki}, R. and {Naylor}, T. and {Oliveira}, J.~M. and {Read}, M. and {Sutorius}, E. and {van Loon}, J. Th. and {Wilkinson}, M.~I. and {Wood}, P.~R.},
        title = "{The VMC survey. I. Strategy and first data}",
      journal = {\aap},
         year = 2011,
        month = mar,
       volume = {527},
          eid = {A116},
        pages = {A116},
          doi = {10.1051/0004-6361/201016137},
archivePrefix = {arXiv},
       eprint = {1012.5193},
 primaryClass = {astro-ph.CO},
       adsurl = {https://ui.adsabs.harvard.edu/abs/2011A&A...527A.116C}
}

@ARTICLE{Zaritsky1997,
       author = {{Zaritsky}, Dennis and {Harris}, Jason and {Thompson}, Ian},
        title = "{A digital photometric survey of the magellanic clouds: First results from one million stars.}",
      journal = {\aj},
         year = 1997,
        month = sep,
       volume = {114},
        pages = {1002-1013},
          doi = {10.1086/118531},
archivePrefix = {arXiv},
       eprint = {astro-ph/9706206},
 primaryClass = {astro-ph},
       adsurl = {https://ui.adsabs.harvard.edu/abs/1997AJ....114.1002Z}
}

@ARTICLE{Abelson2025,
       author = {{Abelson}, C.~S. and {Badenes}, Carles and {Chomiuk}, Laura and {Williams}, Benjamin F. and {Breivik}, Katelyn and {Galbany}, L. and {Jim{\'e}nez-Palau}, C.},
        title = "{The Progenitor Systems of Classical Novae in M31}",
      journal = {\apj},
         year = 2025,
        month = may,
       volume = {984},
       number = {2},
          eid = {134},
        pages = {134},
          doi = {10.3847/1538-4357/adc68c},
archivePrefix = {arXiv},
       eprint = {2501.04925},
 primaryClass = {astro-ph.SR},
       adsurl = {https://ui.adsabs.harvard.edu/abs/2025ApJ...984..134A}
}

@ARTICLE{Sarbadhicary2021,
       author = {{Sarbadhicary}, Sumit K. and {Heiger}, Mairead and {Badenes}, Carles and {Mateu}, Cecilia and {Newman}, Jeffrey A. and {Ciardullo}, Robin and {Hallakoun}, Na'ama and {Maoz}, Dan and {Chomiuk}, Laura},
        title = "{The RR Lyrae Delay-time Distribution: A Novel Perspective on Models of Old Stellar Populations}",
      journal = {\apj},
         year = 2021,
        month = may,
       volume = {912},
       number = {2},
          eid = {140},
        pages = {140},
          doi = {10.3847/1538-4357/abca86},
archivePrefix = {arXiv},
       eprint = {2101.11618},
 primaryClass = {astro-ph.SR},
       adsurl = {https://ui.adsabs.harvard.edu/abs/2021ApJ...912..140S}
}

@ARTICLE{emcee,
       author = {{Foreman-Mackey}, Daniel and {Hogg}, David W. and {Lang}, Dustin and {Goodman}, Jonathan},
        title = "{emcee: The MCMC Hammer}",
      journal = {\pasp},
         year = 2013,
        month = mar,
       volume = {125},
       number = {925},
        pages = {306},
          doi = {10.1086/670067},
archivePrefix = {arXiv},
       eprint = {1202.3665},
 primaryClass = {astro-ph.IM},
       adsurl = {https://ui.adsabs.harvard.edu/abs/2013PASP..125..306F}
}

@ARTICLE{DS21,
       author = {{De Somma}, Giulia and {Marconi}, Marcella and {Cassisi}, Santi and {Ripepi}, Vincenzo and {Pietrinferni}, Adriano and {Molinaro}, Roberto and {Leccia}, Silvio and {Musella}, Ilaria},
        title = "{Period-age-metallicity and period-age-colour-metallicity relations for classical Cepheids: an application to the Gaia EDR3 sample}",
      journal = {\mnras},
         year = 2021,
        month = nov,
       volume = {508},
       number = {1},
        pages = {1473-1488},
          doi = {10.1093/mnras/stab2611},
archivePrefix = {arXiv},
       eprint = {2109.05850},
 primaryClass = {astro-ph.SR},
       adsurl = {https://ui.adsabs.harvard.edu/abs/2021MNRAS.508.1473D}
}

@ARTICLE{Badenes2015,
       author = {{Badenes}, Carles and {Maoz}, Dan and {Ciardullo}, Robin},
        title = "{The Progenitors and Lifetimes of Planetary Nebulae}",
      journal = {\apjl},
         year = 2015,
        month = may,
       volume = {804},
       number = {1},
          eid = {L25},
        pages = {L25},
          doi = {10.1088/2041-8205/804/1/L25},
archivePrefix = {arXiv},
       eprint = {1502.01015},
 primaryClass = {astro-ph.SR},
       adsurl = {https://ui.adsabs.harvard.edu/abs/2015ApJ...804L..25B}
}

@ARTICLE{Riess2019,
       author = {{Riess}, Adam G. and {Casertano}, Stefano and {Yuan}, Wenlong and {Macri}, Lucas M. and {Scolnic}, Dan},
        title = "{Large Magellanic Cloud Cepheid Standards Provide a 1\% Foundation for the Determination of the Hubble Constant and Stronger Evidence for Physics beyond {\ensuremath{\Lambda}}CDM}",
      journal = {\apj},
         year = 2019,
        month = may,
       volume = {876},
       number = {1},
          eid = {85},
        pages = {85},
          doi = {10.3847/1538-4357/ab1422},
archivePrefix = {arXiv},
       eprint = {1903.07603},
 primaryClass = {astro-ph.CO},
       adsurl = {https://ui.adsabs.harvard.edu/abs/2019ApJ...876...85R}
}

@ARTICLE{Hubble1929,
       author = {{Hubble}, E.~P.},
        title = "{A spiral nebula as a stellar system, Messier 31.}",
      journal = {\apj},
         year = 1929,
        month = mar,
       volume = {69},
        pages = {103-158},
          doi = {10.1086/143167},
       adsurl = {https://ui.adsabs.harvard.edu/abs/1929ApJ....69..103H}
}

@ARTICLE{DeSomma2025,
       author = {{De Somma}, Giulia and {Marconi}, Marcella and {Ripepi}, Vincenzo and {Cassisi}, Santi and {Molinaro}, Roberto and {Musella}, Ilaria and {Sicignano}, Teresa and {Trentin}, Erasmo},
        title = "{Spatial Age Distribution of Classical Cepheids in Spiral Galaxies: The Cases of M31 and M33}",
      journal = {\apjl},
         year = 2025,
        month = may,
       volume = {984},
       number = {2},
          eid = {L60},
        pages = {L60},
          doi = {10.3847/2041-8213/adcf92},
archivePrefix = {arXiv},
       eprint = {2504.12638},
 primaryClass = {astro-ph.GA},
       adsurl = {https://ui.adsabs.harvard.edu/abs/2025ApJ...984L..60D}
}

@ARTICLE{Pilecki2022,
       author = {{Pilecki}, Bogumi{\l} and {Thompson}, Ian B. and {Espinoza-Arancibia}, Felipe and {Anderson}, Richard I. and {Gieren}, Wolfgang and {Narloch}, Weronika and {Minniti}, Javier and {Pietrzy{\'n}ski}, Grzegorz and {Taormina}, M{\'o}nica and {Bono}, Giuseppe and {Hajdu}, Gergely},
        title = "{Discovery of a Binary-origin Classical Cepheid in a Binary System with a 59 day Orbital Period}",
      journal = {\apjl},
         year = 2022,
        month = dec,
       volume = {940},
       number = {2},
          eid = {L48},
        pages = {L48},
          doi = {10.3847/2041-8213/ac9fcc},
archivePrefix = {arXiv},
       eprint = {2212.04518},
 primaryClass = {astro-ph.SR},
       adsurl = {https://ui.adsabs.harvard.edu/abs/2022ApJ...940L..48P}
}

@ARTICLE{BohmVitense1998,
       author = {{B{\"o}hm-Vitense}, E. and {Evans}, N.~R. and {Carpenter}, K. and {Albrow}, Michael D. and {Cottrell}, P.~L. and {Robinson}, R. and {Beck-Winchatz}, B.},
        title = "{The Mass of the Cepheid Binary V636 Scorpii}",
      journal = {\apj},
         year = 1998,
        month = oct,
       volume = {505},
       number = {2},
        pages = {903-909},
          doi = {10.1086/306177},
       adsurl = {https://ui.adsabs.harvard.edu/abs/1998ApJ...505..903B}
}

@ARTICLE{Neilson2015,
       author = {{Neilson}, Hilding R. and {Schneider}, Fabian R.~N. and {Izzard}, Robert G. and {Evans}, Nancy R. and {Langer}, Norbert},
        title = "{The occurrence of classical Cepheids in binary systems}",
      journal = {\aap},
         year = 2015,
        month = feb,
       volume = {574},
          eid = {A2},
        pages = {A2},
          doi = {10.1051/0004-6361/201424408},
archivePrefix = {arXiv},
       eprint = {1412.3468},
 primaryClass = {astro-ph.SR},
       adsurl = {https://ui.adsabs.harvard.edu/abs/2015A&A...574A...2N}
}

@ARTICLE{Pietrzynski2010,
       author = {{Pietrzy{\'n}ski}, G. and {Thompson}, I.~B. and {Gieren}, W. and {Graczyk}, D. and {Bono}, G. and {Udalski}, A. and {Soszy{\'n}ski}, I. and {Minniti}, D. and {Pilecki}, B.},
        title = "{The dynamical mass of a classical Cepheid variable star in an eclipsing binary system}",
      journal = {\nat},
         year = 2010,
        month = nov,
       volume = {468},
       number = {7323},
        pages = {542-544},
          doi = {10.1038/nature09598},
archivePrefix = {arXiv},
       eprint = {1012.0231},
 primaryClass = {astro-ph.GA},
       adsurl = {https://ui.adsabs.harvard.edu/abs/2010Natur.468..542P}
}

@ARTICLE{Cuevas-Otahola2025,
       author = {{Cuevas-Otahola}, Bolivia and {Mateu}, Cecilia and {Cabrera-Ziri}, Ivan and {Bruzual}, Gustavo and {Hern{\'a}ndez-P{\'e}rez}, Fabiola and {Magris}, Gladis and {Baumgardt}, Holger},
        title = "{RR Lyrae stars in intermediate-age Magellanic clusters: membership probabilities and delay time distribution}",
      journal = {\mnras},
         year = 2025,
        month = aug,
       volume = {541},
       number = {2},
        pages = {1434-1448},
          doi = {10.1093/mnras/staf1095},
archivePrefix = {arXiv},
       eprint = {2411.12741},
 primaryClass = {astro-ph.GA},
       adsurl = {https://ui.adsabs.harvard.edu/abs/2025MNRAS.541.1434C}
}

@ARTICLE{Mateu2025,
       author = {{Mateu}, Cecilia and {Cuevas-Otahola}, Bolivia and {Jos{\'e} Downes}, Juan},
        title = "{First direct detection of an RR Lyrae star conclusively associated with an intermediate-age cluster}",
      journal = {arXiv e-prints},
         year = 2025,
        month = sep,
          eid = {arXiv:2509.22336},
        pages = {arXiv:2509.22336},
          doi = {10.48550/arXiv.2509.22336},
archivePrefix = {arXiv},
       eprint = {2509.22336},
 primaryClass = {astro-ph.SR},
       adsurl = {https://ui.adsabs.harvard.edu/abs/2025arXiv250922336M}
}

@ARTICLE{Iorio2021,
       author = {{Iorio}, Giuliano and {Belokurov}, Vasily},
        title = "{Chemo-kinematics of the Gaia RR Lyrae: the halo and the disc}",
      journal = {\mnras},
         year = 2021,
        month = apr,
       volume = {502},
       number = {4},
        pages = {5686-5710},
          doi = {10.1093/mnras/stab005},
archivePrefix = {arXiv},
       eprint = {2008.02280},
 primaryClass = {astro-ph.GA},
       adsurl = {https://ui.adsabs.harvard.edu/abs/2021MNRAS.502.5686I}
}

@ARTICLE{Marconi2013,
       author = {{Marconi}, M. and {Molinaro}, R. and {Ripepi}, V. and {Musella}, I. and {Brocato}, E.},
        title = "{Theoretical fit of Cepheid light a radial velocity curves in the Large Magellanic Cloud cluster NGC 1866}",
      journal = {\mnras},
         year = 2013,
        month = jan,
       volume = {428},
       number = {3},
        pages = {2185-2197},
          doi = {10.1093/mnras/sts197},
archivePrefix = {arXiv},
       eprint = {1210.4343},
 primaryClass = {astro-ph.SR},
       adsurl = {https://ui.adsabs.harvard.edu/abs/2013MNRAS.428.2185M}
}

@ARTICLE{Ragosta2019,
       author = {{Ragosta}, Fabio and {Marconi}, Marcella and {Molinaro}, Roberto and {Ripepi}, Vincenzo and {Cioni}, Maria Rosa L. and {Moretti}, Maria Ida and {Groenewegen}, Martin A.~T. and {Choudhury}, Samyaday and {de Grijs}, Richard and {van Loon}, Jacco Th and {Oliveira}, Joana M. and {Ivanov}, Valentin D. and {Gonzalez-Fernandez}, Carlos},
        title = "{The VMC survey - XXXV. model fitting of LMC Cepheid light curves}",
      journal = {\mnras},
         year = 2019,
        month = dec,
       volume = {490},
       number = {4},
        pages = {4975-4984},
          doi = {10.1093/mnras/stz2881},
archivePrefix = {arXiv},
       eprint = {1910.05052},
 primaryClass = {astro-ph.SR},
       adsurl = {https://ui.adsabs.harvard.edu/abs/2019MNRAS.490.4975R}
}

@ARTICLE{Magnier1997,
       author = {{Magnier}, E.~A. and {Augusteijn}, T. and {Prins}, S. and {van Paradijs}, J. and {Lewin}, W.~H.~G.},
        title = "{Cepheids as tracers of star formation in M 31. I. Observations and identifications}",
      journal = {\aaps},
         year = 1997,
        month = dec,
       volume = {126},
        pages = {401-406},
          doi = {10.1051/aas:1997394},
       adsurl = {https://ui.adsabs.harvard.edu/abs/1997A&AS..126..401M}
}

\end{document}